\documentclass[11pt]{article}

\usepackage{graphicx}
\usepackage{amsmath}
\usepackage{dcolumn}
\usepackage{bm}
\usepackage{slashed}
\usepackage[bookmarks,bookmarksnumbered,colorlinks=true,anchorcolor=blue,
linkcolor=blue,urlcolor=blue,citecolor=blue,breaklinks=true]{hyperref}
\usepackage[utf8]{inputenc}
\usepackage{authblk}
\usepackage{cite}
\usepackage{titlesec}
\usepackage{caption}
\usepackage{subcaption}

\usepackage{color}
\usepackage{ulem}

\newcommand{\be}{\begin{equation}}
\newcommand{\ee}{\end{equation}}
\newcommand{\ba}{\begin{eqnarray}}
\newcommand{\ea}{\end{eqnarray}}
\def\bea{\begin{eqnarray}}
\def\eea{\end{eqnarray}}

\newcommand{\gsim}{\mathrel{\hbox{\rlap{\lower.55ex \hbox {$\sim$}}
                   \kern-.3em \raise.4ex \hbox{$>$}}}}
\newcommand{\lsim}{\mathrel{\hbox{\rlap{\lower.55ex \hbox {$\sim$}}
                   \kern-.3em \raise.4ex \hbox{$<$}}}}

\def\roughly#1{\mathrel{\raise.3ex\hbox{$#1$\kern-.75em%
\lower1ex\hbox{$\sim$}}}}
\def\lsim{\roughly<}
\def\gsim{\roughly>}

\def\({\left(}
\def\){\right)}
\def\[{\left[}
\def\]{\right]}
\def\<{\langle}
\def\>{\rangle}

\usepackage{xcolor}

\begin{document}
\title{\bf The thread embodiment of holographic quantum entanglement}

\author[]{Yi-Yu Lin$^{1,2}$ \thanks{yiyu@simis.cn}}

 \affil{${}^2$Fudan Center for Mathematics and Interdisciplinary Study, Fudan University, Shanghai, 200433, China}
\affil[]{${}^1$Shanghai Institute for Mathematics and Interdisciplinary Sciences (SIMIS), Shanghai, 200433, China}


\maketitle

\begin{abstract}

This paper systematically develops the concept of entanglement threads that characterize the entanglement structure of holographic duality. Behind this framework lies a simple philosophy: holographic quantum entanglement can be visualized using thread-like objects. Inspired by the fact that tensor network models can be deformed into a quantum circuit form with flow-conserving features, we abstract the concept of entanglement threads. These entanglement threads can be understood as a pre-set ensemble of wires in a holographic quantum circuit, and we propose that they characterize the underlying partially ordered structure of holographic quantum entanglement. Combining the concepts of entanglement threads and kinematic space, a elegant circuit interpretation for the holographic complexity is provided. We also clarify the connection and distinction between entanglement threads and the previously proposed concept of bit threads.

\end{abstract}
\tableofcontents

\newpage

\section{Introduction}

Many phenomena in the holographic principle~\cite{Maldacena:1997re,Gubser:1998bc,Witten:1998qj}, especially the holographic entanglement entropy formula proposed in 2006 (the Ryu-Takayanagi formula) ~\cite{Ryu:2006bv,Ryu:2006ef,Hubeny:2007xt}, reveal a close connection between gravitational spacetime geometry and quantum entanglement. Along the line of thought that spacetime geometry may be constructed from quantum entanglement, an obviously important direction is to fully understand the nature of the entanglement structure that corresponds to classical spacetimes dual to holographic quantum systems in AdS/CFT duality. In 2009, Swingle was the first to propose that tensor networks, developed in the study of quantum many-body systems, could be used to characterize this entanglement structure~\cite{Swingle:2009bg, Swingle:2012wq}. Tensor networks are widely used in quantum many-body theory as a powerful numerical simulation tool, representing a quantum state as a contraction of many smaller tensors. Each small tensor can be regarded as a local few-body quantum state and can be represented diagrammatically as a vertex from which several legs extend, corresponding to local qudits. With such a diagrammatic representation, the whole quantum state is naturally depicted as a network with geometric structure. The MERA (Multiscale Entanglement Renormalization Ansatz) tensor network~\cite{Vidal:2007hda,Vidal:2008zz,Vidal:2015}, a tensor network specifically designed to represent the ground states of critical systems (which can be viewed as lattice versions of conformal field theories), exhibits a high-dimensional geometric structure strikingly similar to a spatial slice of AdS space. Moreover, in the MERA tensor network, the minimal cut that divides the network into two parts (i.e., the one that cuts through the minimal number of internal legs) provides an estimate of the entanglement entropy of a subregion. This strongly resembles the RT formula. Therefore, the MERA tensor network is regarded as a concrete realization of the AdS/CFT duality. Due to the physical meaning of tensor networks, it is natural that the geometric structure of the MERA tensor network encodes information about quantum entanglement. The approach of using tensor networks to study the entanglement structure in holographic duality has been continuously developed, and many other concrete models of holographic tensor networks have been proposed and improved, see e.g. ~\cite{ Evenbly:2017hyg,Milsted:2018vop, Milsted:2018yur, Milsted:2018san, McMahon:2018okk, Pastawski:2015qua , Nezami:2016zni, Hayden:2016cfa,Yang:2015uoa,Qi:2018shh, Jahn:2019nmz, Bhattacharyya:2017aly,Hung:2019zsk,Chen:2021ipv, Chen:2021rsy, Chen:2021qah, Hung:2024gma, Geng:2025efs, Chen:2022wvy, Chirco:2017vhs, Han:2016xmb, Colafranceschi:2021acz, Colafranceschi:2022ual, Singh:2017tet, Singh:2017xjx, Bao:2018pvs, Lin:2020ufd}.

Another novel approach to studying the entanglement structure in holographic duality is based on a ``thread picture" (as opposed to the ``network picture" of tensor networks). This thread-based picture first attracted attention in 2016, when Headrick and collaborators developed the concept of bit threads~\cite{Freedman:2016zud,Cui:2018dyq,Headrick:2017ucz,Headrick:2022nbe}. Consider partitioning a holographic quantum system into two parts, $A$ and $\bar{A}$. A bit thread is a one-dimensional line that starts from region $A$, traverses the holographic bulk, and connects to region $\bar{A}$. By drawing an analogy with the max-cut/min-flow theorem in network flow theory~\cite{mincut1, mincut2}, they discovered that if the density of bit threads in the bulk is constrained by an upper bound, then the maximum possible number of threads connecting $A$ and $\bar{A}$ is exactly given by the area of the Ryu-Takayanagi surface—a minimal extremal surface. The bit thread bundle vividly characterizes the entanglement entropy between $A$ and $\bar{A}$, as the RT surface area equals the entanglement entropy according to the RT formula. Furthermore, this naturally connects to the concept of entanglement distillation \cite{Bennett:1995tk, Bennett:1995ra, Bennett:1996gf} in quantum information theory.

In this paper, inspired by the thread picture that naturally emerges in holographic tensor network models, we abstract the notion of entanglement threads. These entanglement threads traverse codimension-one spatial slices of the holographic bulk, with their endpoints anchored on the holographic boundary. We argue that the number of entanglement threads connecting two subregions on the holographic boundary is given by the half-conditional mutual information. Furthermore, we demonstrate that the specific trajectory of each entanglement thread in the bulk is precisely a geodesic. This leads us to view the kinematic space~\cite{Czech:2015kbp,Czech:2015qta} as a ``circuit board'' that organizes qudits, with entanglement threads serving as wires extending from it, utilized to construct quantum circuits for the corresponding holographic states. Therefore, the collection of entanglement threads provides a pre-structure for adding basic tensors or quantum gates. In this sense, we propose that entanglement threads encode some underlying partial order information of the entanglement structure in holographic duality. We also show that each entanglement thread can be assigned with an interpretation of thread–state correspondence, characterizing how the qudits it threads are globally entangled.  Ultimately, the entanglement relationships are intuitively captured—not only between various subregions on the holographic boundary, but also between different surfaces within the bulk, via entanglement threads that simultaneously pass through them.

The structure of this paper is as follows. In Section~\ref{sec1}, we introduce the background and motivation, presenting the basic setup of the entanglement thread picture along with the inspiration from tensor networks. In Section~\ref{sec2}, we clarify the rules governing the trajectories of entanglement threads in the holographic bulk, helping us confirm that their paths in the bulk slices are exactly geodesics. In Section~\ref{sec3}, we introduce the mathematical tools of kinematic space, which are particularly well-suited for organizing these entanglement threads. We demonstrate how entanglement threads can be used as wires to construct a quantum circuit representing holographic quantum states, and verify this idea from the perspective of holographic complexity. In Section~\ref{sec4}, we further discuss the physical significance of entanglement threads and their connections and distinctions from the earlier bit thread concept. Section~\ref{sec5} is the conclusion. For completeness, a more detailed review of kinematic space is provided in the Appendix.


\section{Background and Motivation}\label{sec1}
\subsection{Representing Holographic Quantum Entanglement as Threads}\label{sec11}

\begin{figure}\label{figure1}
     \centering
     \begin{subfigure}[b]{0.38\textwidth}
         \centering
         \includegraphics[width=\textwidth]{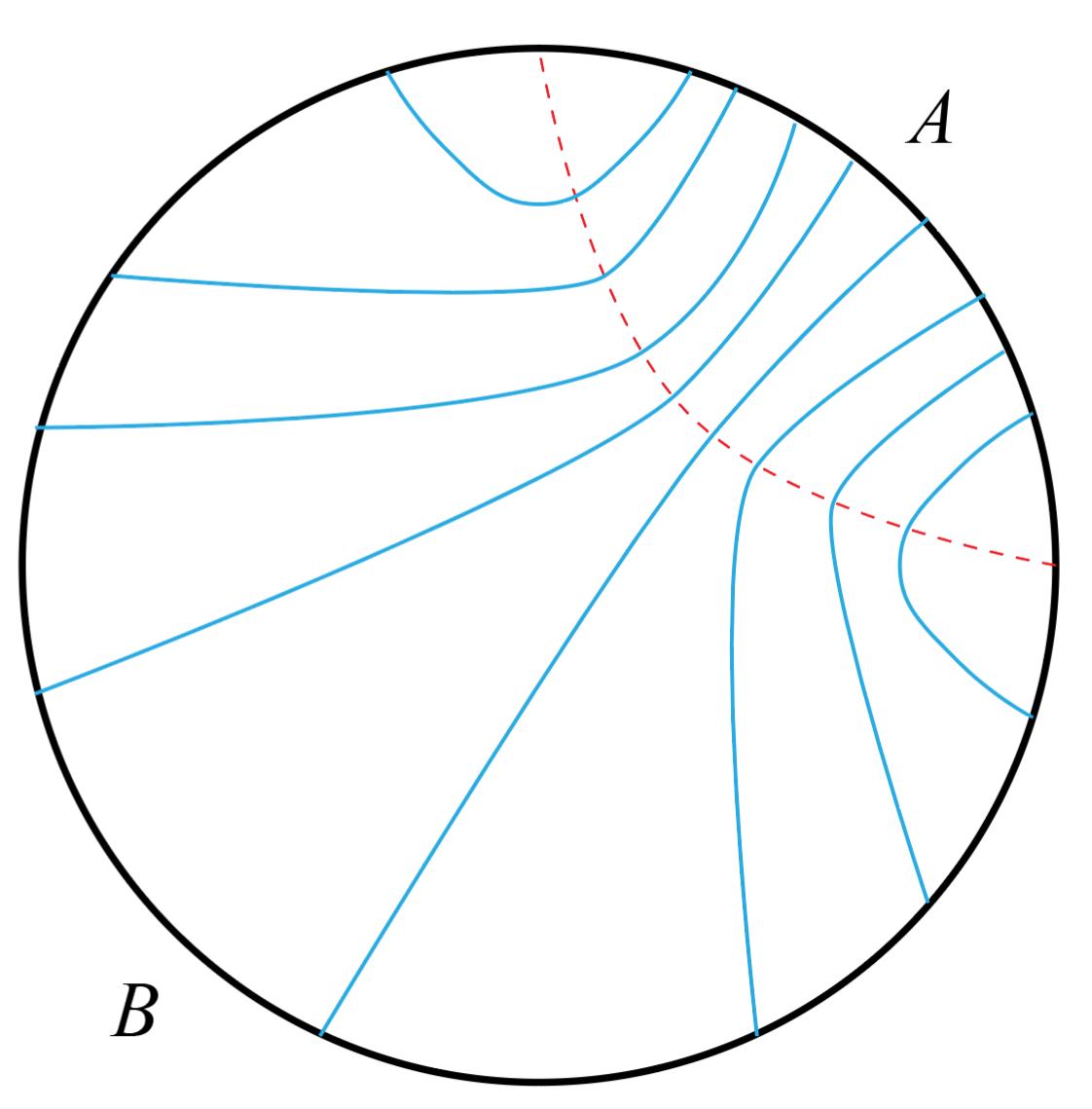}
         \caption{}
         \label{2.1.1a}
     \end{subfigure}
     \hfill
     \begin{subfigure}[b]{0.4\textwidth}
         \centering
         \includegraphics[width=\textwidth]{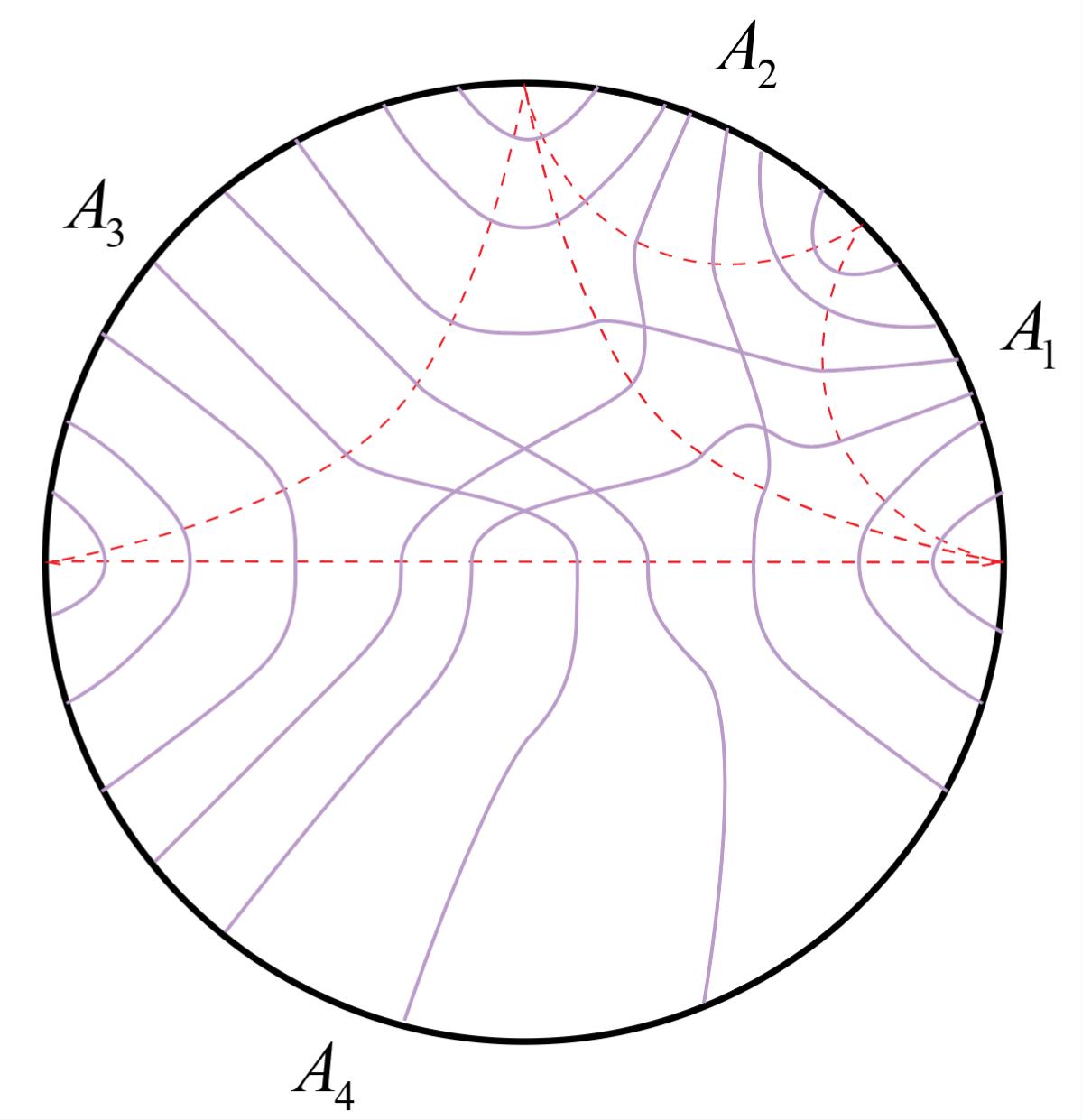}
         \caption{}
         \label{2.1.1b}
     \end{subfigure}
     \hfill
     \begin{subfigure}[b]{1\textwidth}
         \centering
         \includegraphics[width=\textwidth]{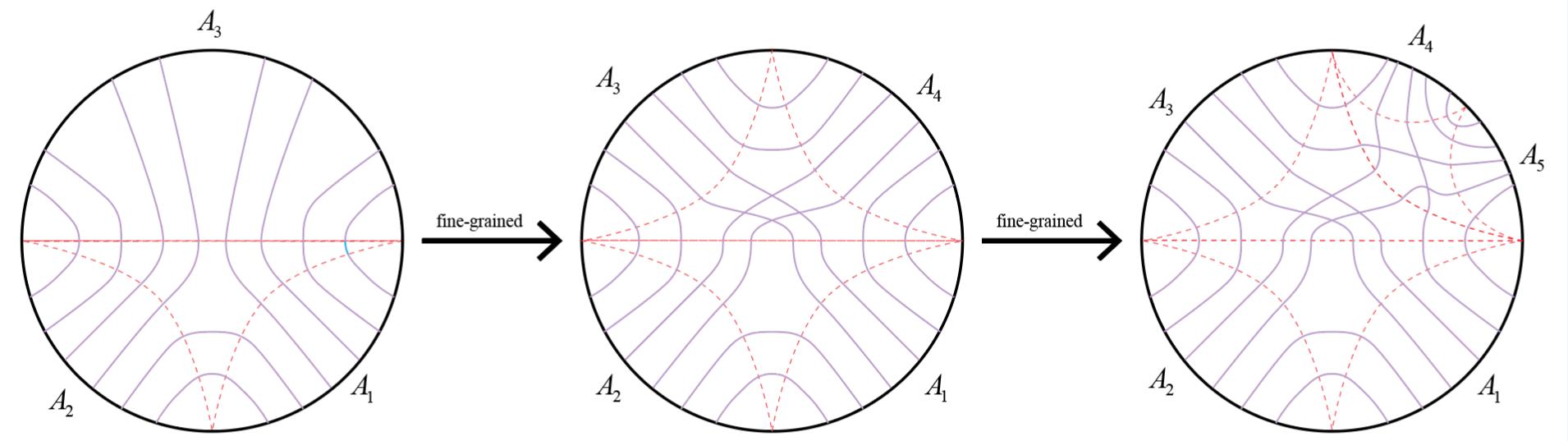}
         \caption{}
         \label{refine}
     \end{subfigure}
     \caption{(a) A holographic ``thread'' picture characterizing the entanglement entropy between two complementary regions. (b) A more refined thread configuration characterizing a set of entanglement entropies involving more subregions. (c) By iteratively dividing the quantum system, more and more refined thread configurations can be constructed, characterizing the entanglement structure at more and more refined levels.  Here the RT surfaces are represented as red dashed lines, while the threads are schematically represented as purple lines. The pictures here are basically  taken from another paper~\cite{Lin:2023jah}. However, please note that the exact trajectories of these entanglement threads have not yet been really specified, which is precisely what this present paper aims to achieve.}
\end{figure}

The idea of representing holographic quantum entanglement using threads is, in a sense, simple and straightforward\footnote{Motivated by the concept of bit threads~\cite{Freedman:2016zud,Cui:2018dyq,Headrick:2017ucz,Headrick:2022nbe}, the  thread-kind objects have been used in a series of works
(see e.g.,~\cite{Lin:2023jah,Lin:2022flo,Lin:2022agc,Lin:2020yzf,Lin:2023orb,Lin:2021hqs, Jahn:2019nmz, Yan:2019vzp,Yang:2015uoa,Lin:2024dho,Lin:2023rxc}) to attempt to describe the connection between quantum entanglement and spacetime structure. Despite of their different perspectives, these works essentially express a rather natural idea to decompose entanglement entropy in a more refined manner. This paper can be regarded as a unified description of these thread objects, by abstracting the concept of entanglement threads from the general tensor network model.}
. Since entanglement entropy measures the degree of quantum entanglement, for example, if we consider two complementary systems $ A $ and $ B $, and wish to depict a contribution of one unit ($ \log 2 $) of entanglement to the entanglement entropy, we may schematically draw a line (``thread”) connecting $ A $ and $ B $. Thus, when the entanglement entropy between $ A $ and $ B $ is $ n \log 2 $, we draw $ n $ threads between $ A $ and $ B $. Generalizing this to multiple subsystems $ A_1, A_2, A_3, \dots, A_N $ is straightforward; we simply imagine each pair $ A_i $ and $ A_j $ being connected by a bundle of threads, whose number is denoted by $ F_{ij} $. It is proven in \cite{Lin:2021hqs} that in order to correctly match the entanglement entropy for any bipartition of the total system (which we assume to be in a pure state), the number of threads $ F_{ij} $ connecting regions $ A_i $ and $ A_j $ is precisely given by one half of the quantum conditional mutual information (qCMI) (as we will review in the following).

In the holographic framework, one observes that in addition to the quantum system itself, there naturally exists a higher-dimensional bulk. This makes the idea of characterizing entanglement via threads even more intuitive! From now on, we shall appropriately refer to these threads as holographic entanglement threads, or simply entanglement threads. We can imagine these threads being carried by—or threading through—codimension-1 spatial slices of the bulk, as originally inspired by the bit threads proposed in ~\cite{Freedman:2016zud,Cui:2018dyq,Headrick:2017ucz,Headrick:2022nbe}. In particular, this idea aligns closely with the Ryu–Takayanagi (RT) proposal for holographic entanglement entropy ~\cite{Ryu:2006bv,Ryu:2006ef,Hubeny:2007xt}. To see this, consider the simplest case of a subregion $ A $ in a CFT pure state that is dual to a pure AdS bulk. We may imagine a family of uninterrupted threads, each connecting $ A $ and its complement $ B $, passing through the RT surface $ \gamma_A $, with the number of threads equal to the entanglement entropy between $ A $ and $ B $, denoted by $ S(A) $ (see Figure~\ref{2.1.1a}):
\begin{equation}
F_{AB} = S_A.
\end{equation}
By the RT formula~\cite{Ryu:2006bv,Ryu:2006ef,Hubeny:2007xt} (setting $ 4G_N = 1 $), the area of $ \gamma_A $ equals $ S_A $, so we further have:
\begin{equation}
F_{AB} = \text{Area}(\gamma_A),
\end{equation}
which means that each unit area of $ \gamma_A $ accommodates exactly one entanglement thread.

\begin{figure}
    \centering
    \includegraphics[scale=0.3]{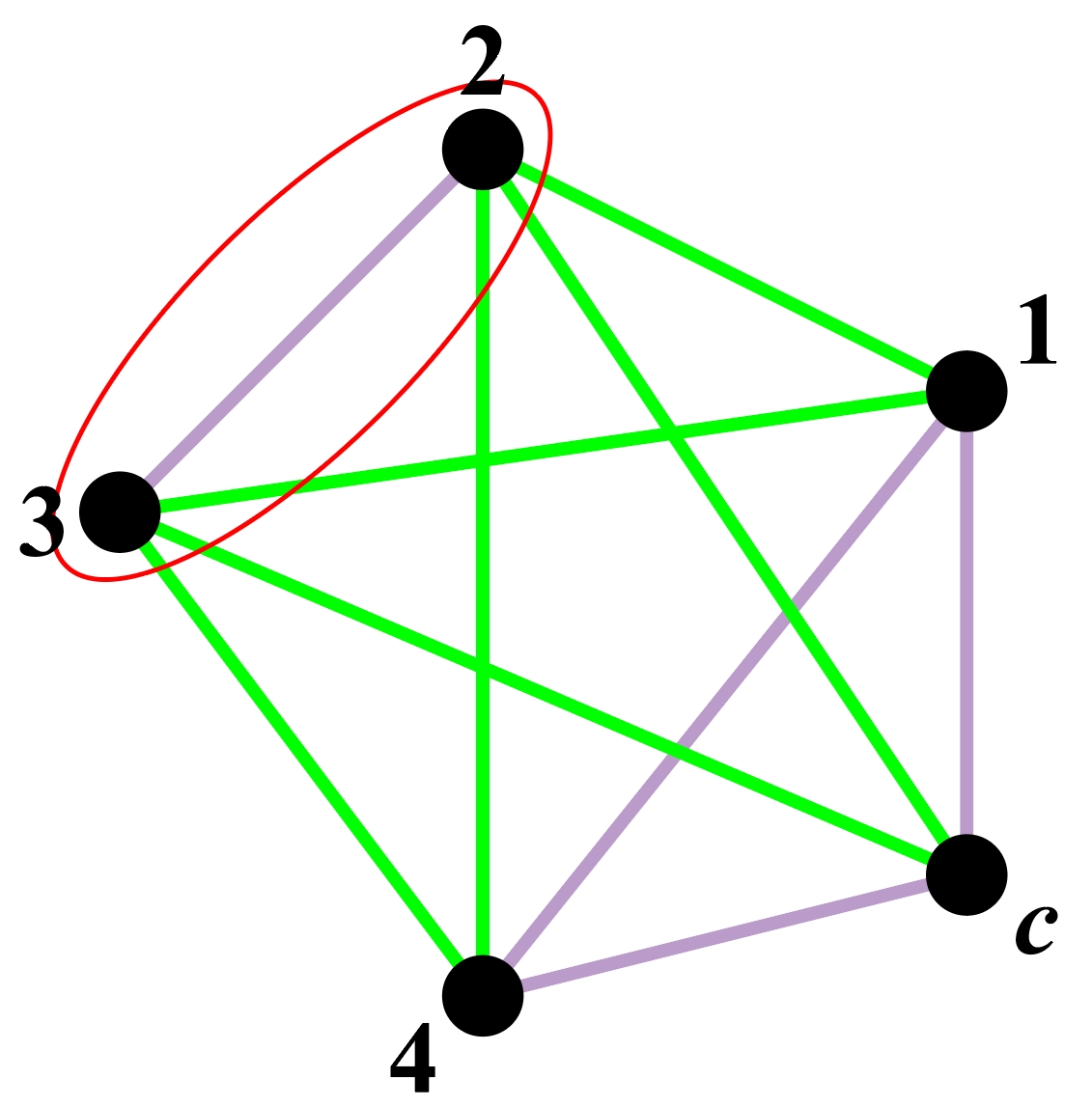}
    \caption{A simplified network diagram corresponding to the entanglement thread configuration. Each elementary region is simplified by a point and the component flow flux $ F_{ij} $ is represented by the line segment connecting two points. Then for example, the entropy of the region in red circle is equal to the sum of the green lines in this figure. }    \label{21.2}
\end{figure}

As mentioned earlier, we can go beyond the bipartite case and do better. We can construct increasingly refined thread configurations that allow us to compute the entanglement entropies for not just a single subregion. For instance, as shown in Figure~\ref{2.1.1b}, we can further decompose region $ A $ into $ A = A_1 \cup A_2 $, and $ B $ into $ B = A_3 \cup A_4 $. Thus, we can construct a finer thread configuration that enables us to compute the entanglement entropies between six connected regions ($ A_1, A_2, A_3, A_4 $, as well as $ A = A_1 \cup A_2 $ and $ B = A_3 \cup A_4 $) and their complements. Specifically, we design the entanglement thread configuration so that the number of threads connecting each of these six regions to their complements matches the corresponding entanglement entropy. Again, using the RT formula, we can describe that on each of the six connected RT surfaces $ \gamma_1, \gamma_2, \gamma_3, \gamma_4, \gamma_{12}, \gamma_{34} $, there is exactly one entanglement thread per unit area. This ``thought experiment'' can be iterated: we further partition the quantum system $ M $ into more adjacent and non-overlapping ``elementary regions'' $ A_1, A_2, \dots, A_N $~\footnote{ We can take $ N $ to be very large, expecting to capture more refined entanglement structure.}, with:
\begin{equation}
A_i \cap A_j = \emptyset, \quad \bigcup_i A_i = M,
\end{equation}
and obtain a correspondingly finer thread configuration (see Figure~\ref{refine}). It is preferable that each elementary region remains much larger than the Planck scale, to carefully ensure the applicability of the RT formula. Let us denote the number of entanglement threads between each pair of regions $ A_i $ and $ A_j $ as $ F_{ij} $. Following~ \cite{Lin:2021hqs, Lin:2023orb}, we describe this thread configuration using a complete graph, as shown in Figure~\ref{21.2}, where each region $ A_i $ is simplified to a boundary vertex and each thread bundle between regions is simplified to an edge connecting two vertices. The value $ F_{ij} $ is represented by the weight of the edge. Adjacent elementary regions can be merged into simply connected regions, e.g., $ A_{12} = A_1 \cup A_2 $, and all such simply connected unions form a set: $\{ A_{i(i+1)\cdots j} \mid i \le j \}$, whose number of elements is $\frac{N(N-1)}{2}$. Each simply connected region $ A_{i(i+1)\cdots j} $ has an entanglement entropy $ S_{i(i+1)\cdots j} $ with its complement (with the assumption that $ M $ is in a pure state). According to the thread representation of entanglement, we write:
\begin{equation}
S_{a(a+1)\cdots b} = \sum_{i \in \{a,\dots,b\},\; j \notin \{a,\dots,b\}} F_{ij}, \label{equ}
\end{equation}
This can be intuitively understood as follows: the entanglement entropy between $ A = A_{i(i+1)\cdots j} $ and its complement $ \bar{A} $ comes from the total number of threads connecting elementary regions within $ A $ and those in $ \bar{A} $. Given the physical meaning of $ F_{ij} $ as the number of threads connecting $ A_i $ and $ A_j $, the values of $ F_{ij} $ are evidently symmetric in $ i $ and $ j $, so their total number is also $ \frac{N(N-1)}{2} $, matching the number of entropies $ S_{i(i+1)\cdots j} $. Thus, the system of equations \eqref{equ} is full rank and has a unique solution \cite{Lin:2021hqs}, given by:
\begin{equation}
F_{ij} = \frac{1}{2} I(A_i, A_j \mid L) = \frac{1}{2} \left[ S(A_i \cup L) + S(A_j \cup L) - S(L) - S(A_i \cup L \cup A_j) \right], \label{half}
\end{equation}
where $ L = A_{(i+1)\cdots(j-1)} $ denotes the simply connected region between $ A_i $ and $ A_j $, encoding the size of their separation. In other words, the number of threads connecting $ A_i $ and $ A_j $ is precisely given by the half-conditional mutual information.

At this point, we should note that in using the simplified complete graph and Eq.~\eqref{equ}, we have so far assigned these threads only a topological meaning. That is, although we believe the higher-dimensional bulk hosts these threads, we have not specified their actual trajectories in the bulk. In fact, the goal of this paper is precisely to explore: 

\textit{To what extent is this idea of representing CMI as threads reasonable, and how should we understand these threads, especially their trajectories in the holographic bulk...}

It is also worth emphasizing that CMI does not have to be visualized as thread-like objects. This approach is motivated by several reasons. First, it is a naive attempt to depict entanglement between two subsystems by connecting them with threads—the more entangled, the more threads. Second, the idea is inspired by bit threads ~\cite{Freedman:2016zud, Cui:2018dyq, Headrick:2017ucz}, where entanglement entropy can be encoded in the number of threads. Third, in the special context of holography, the extra bulk dimension naturally serves to accommodate the trajectories of these threads.

Finally, as background, we note that conditional mutual information appears naturally in various studies of holographic duality, such as in holographic kinematic space ~\cite{Czech:2015kbp,Czech:2015qta}, the holographic entropy cone ~\cite{Bao:2015bfa,Hubeny:2018ijt,Hubeny:2018trv,HernandezCuenca:2019wgh}, and holographic partial entanglement entropy ~\cite{Vidal:2014aal,Wen:2019iyq,Wen:2018whg,Kudler-Flam:2019oru}. The reason is quite understandable: these studies all focus on exploring possible decompositions of holographic entanglement entropy, and the ``visual metaphor” we just discussed reveals a very simple underlying essence.

Next, we will soon turn to the topic of tensor networks. After gaining some inspiration there, we will return to the topic of threads.


\subsection{The Thread Picture in Tensor Network Models}\label{sec12}

\begin{figure}
     \centering
       \begin{subfigure}[b]{0.4\textwidth}
         \centering
         \includegraphics[width=\textwidth]{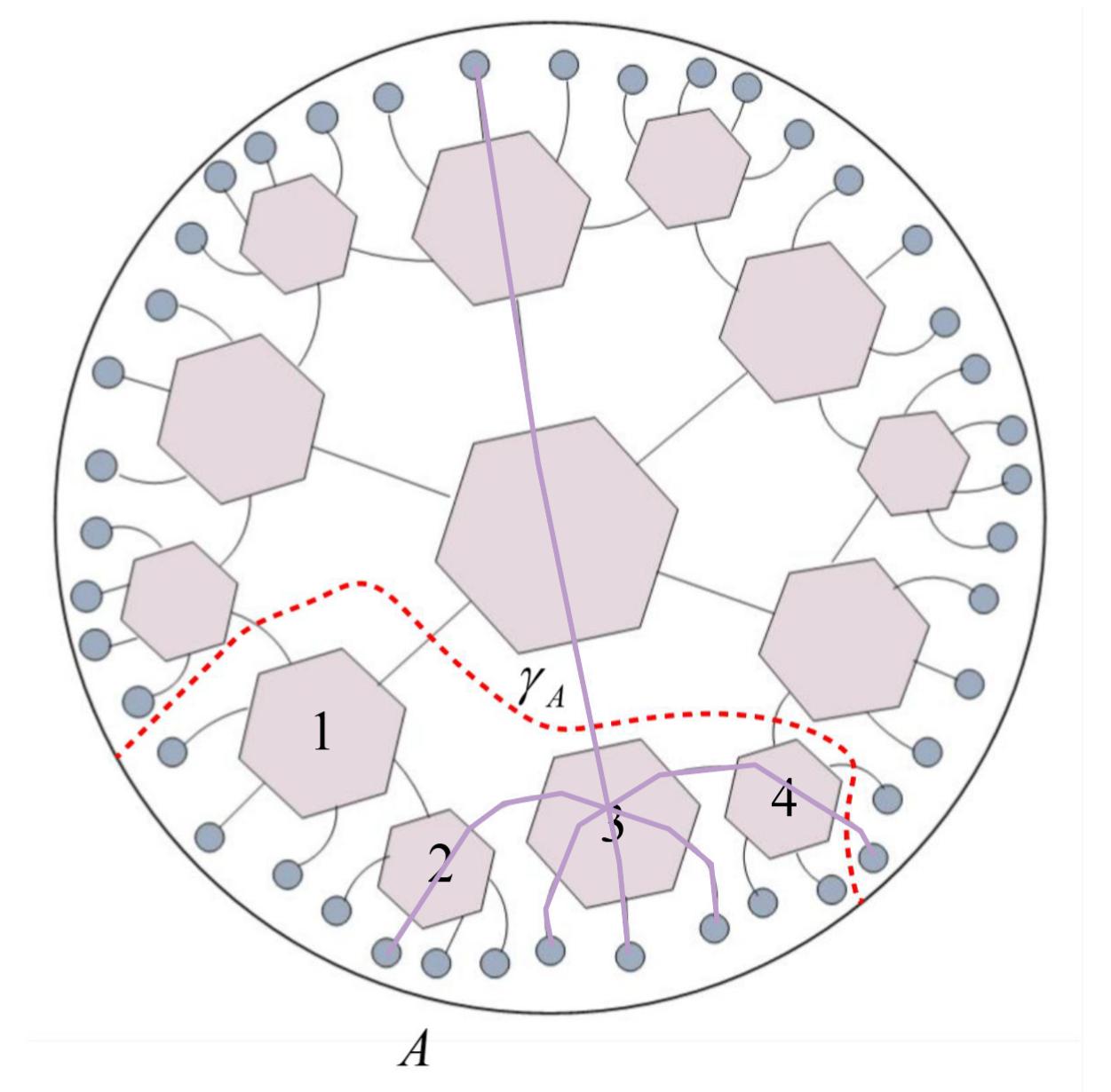}
         \caption{}
         \label{22.1a}
     \end{subfigure}
     \hfill
     \begin{subfigure}[b]{0.4\textwidth}
         \centering
         \includegraphics[width=\textwidth]{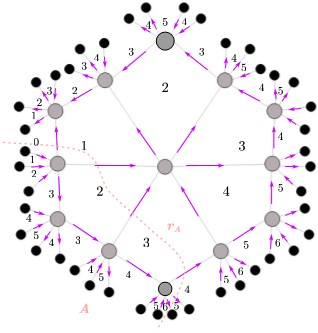}
         \caption{}
         \label{22.1b}
     \end{subfigure}
\hfill
     \begin{subfigure}[b]{0.4\textwidth}
         \centering
         \includegraphics[width=\textwidth]{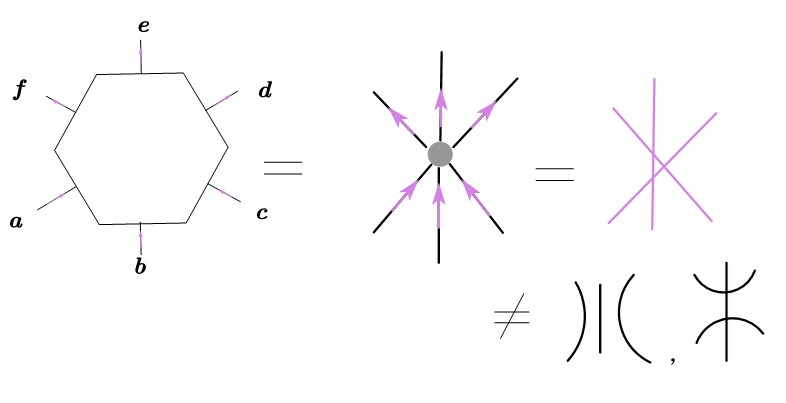}
         \caption{}
         \label{22.1c}
     \end{subfigure}
\hfill
     \begin{subfigure}[b]{0.4\textwidth}
         \centering
         \includegraphics[width=\textwidth]{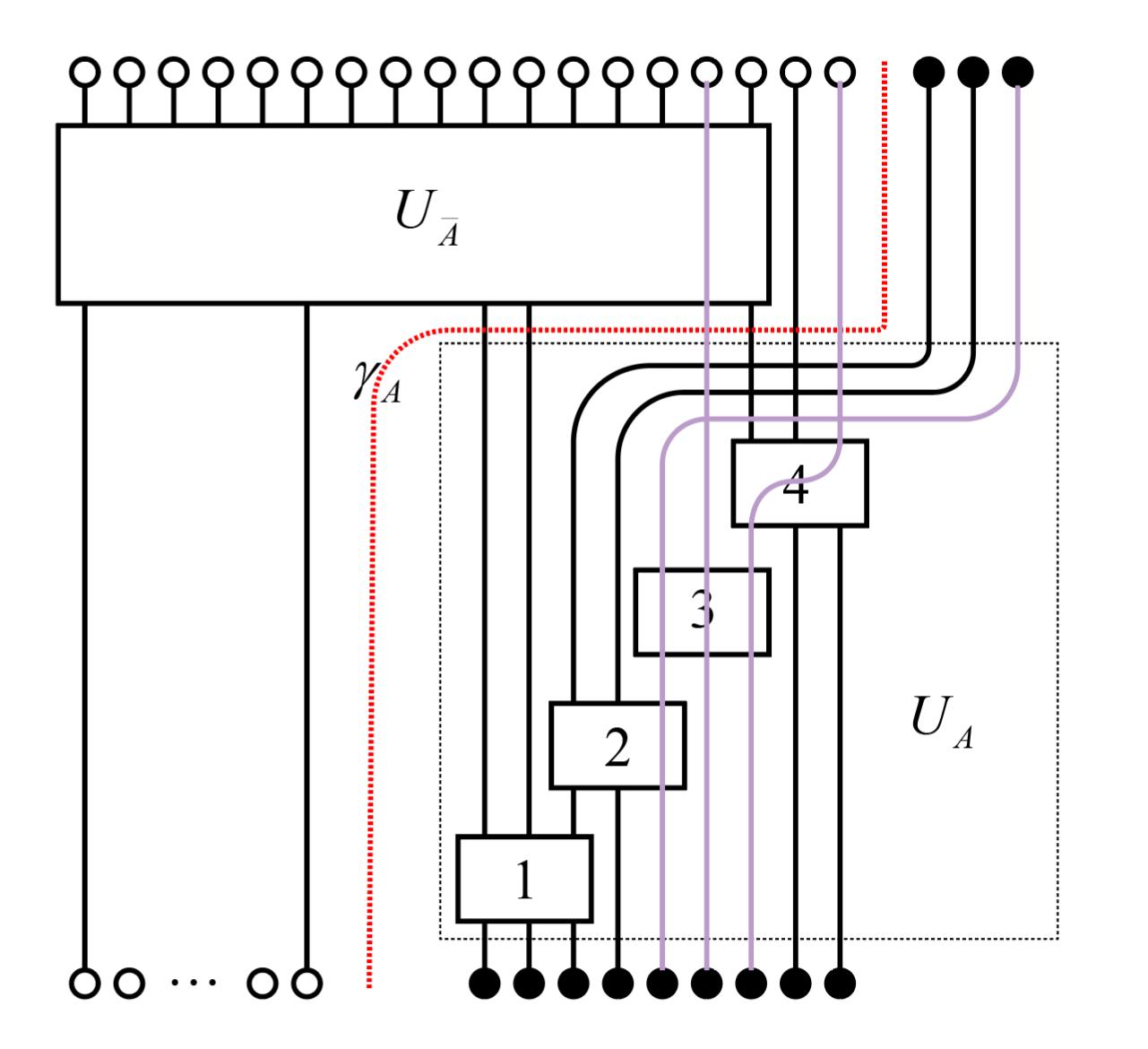}
         \caption{}
\label{22.1d}\end{subfigure}\caption{(a) HaPPY tensor network state and an RT surface (red dashed line) which calculates the entanglement entropy of a subregion $A$. Furthermore, in Section~\ref{sec23}, we will obtain the specific trajectory of the entanglement thread (the purple line) therein. (b) Following~\cite{Pastawski:2015qua}, we assign orientations to the edges (i.e., legs) of the HaPPY network. The global ``flow'' structure of the entire network naturally generates a picture similar to the thread configuration--by gluing together many small arrows pointing in the same direction to form a single thread. (c) In principle, there is more than one way to connect arrows of the same direction into thread-like objects. However, the rules in Section~\ref{sec22} fully determine the concrete configuration of entanglement threads.  (d) Deforming the tensor network topologically into a quantum circuit. Then an entanglement thread (the purple line) corresponds to a quantum wire therein. In particular, we have demonstrated three entanglement threads as explanatory aids (for both (a) and (d)). }\label{22.1}\end{figure}

In this section, we briefly shift to the topic of tensor networks, as a review of this will provide us with significant insights into the thread picture. We will realize that in tensor network models of holographic duality, structures resembling the entanglement thread picture described above emerge naturally and ubiquitously. Beyond this phenomenological inspiration, the thread picture within tensor networks also endows entanglement threads with more precise quantum-state interpretations and helps us understand the meaning of their trajectories in the bulk.

We take the example of the ``HaPPY tensor network state”. Based on the close connection between holographic states and quantum error-correcting codes~\cite{Almheiri:2014lwa,Dong:2016eik}, ~\cite{Pastawski:2015qua} proposed a tensor network composed of perfect tensors, as shown in Figure \ref{22.1a}, to describe the quantum states of a holographic quantum system. Each hexagonal basic unit extends six legs, representing a basic perfect tensor of six qubits $T_{abcdef}$. A $2s$-perfect tensor is a pure state of $2s$ qudits such that, for any positive integer $k \le s$, the mapping from any $k$ qudits to the remaining $2s-k$ qudits is an isometry~\cite{Facchi:2008ora, Helwig:2012nha , Helwig:2013ckb, Helwig:2013qoq }. 

It is noteworthy that the original paper~\cite{Pastawski:2015qua} has already revealed the presence of an evident thread picture in the HaPPY tensor network. More specifically, as pointed out in~\cite{Pastawski:2015qua}, to calculate the entanglement entropy of a connected region $A$ within the framework of the HaPPY tensor network, one can perform a topological deformation of the tensor network into a quantum circuit. As shown in Figure \ref{22.1d}, after deforming the tensor network topologically into a quantum circuit, the entanglement entropy of $A$ is simply given by the number of wires that pass through the RT surface $\gamma_A$. For completeness, we now briefly review the method of transforming the HaPPY tensor network into a quantum circuit. Following~\cite{Pastawski:2015qua}, we first assign orientations to the edges (i.e., legs) of the HaPPY network. As shown in Figure~\ref{22.1b} (where, we represent the hexagonal pattern simply as a circular vertex), this can be achieved by assigning numbers to the nodes of the dual graph.~\footnote{ In graph theory, a node in the dual graph corresponds to a region enclosed by edges in the original network (at the boundary, this is given by the gaps between the external legs).} The procedure is as follows: first, choose any boundary node (i.e., the ``gap” between two external legs) as the ``starting point" and label it with the number 0. Then, assign a number to each dual node such that the number equals the distance between that node and the starting node. In other words, in the original network, the number assigned to a region is the minimum number of edges that need to be crossed to reach it from region 0. These numbers then determine the orientation of each edge in the HaPPY network, according to the following rule: the number on the right side of an ``arrow” (i.e., an directed edge) is always greater than that on the left side. As shown in Figure~\ref{22.1b}, the global ``flow'' structure of the entire network naturally generates a picture similar to the thread configuration--the idea is simply to glue together many small arrows pointing in the same direction to form a single thread. As pointed out in~\cite{Pastawski:2015qua}, this is facilitated by the fact that each perfect tensor has an even number of legs, ensuring that the number of incoming arrows at each vertex in the network equals the number of outgoing arrows. This guarantees ``flow conservation", which makes defining a thread configuration possible. Moreover, the negatively curved tiling pattern of the HaPPY model ensures the uniqueness of edge directionality and the absence of circular flows, which matches the default rule of entanglement threads: each thread connects two boundary points. Using the direction of arrows in the network as a partial order, we can topologically deform the directed HaPPY tensor network graph into a quantum circuit diagram, as shown in Figure~\ref{22.1d}, where each basic tensor is interpreted as a quantum gate with three inputs and three outputs.

Based on the above observations and insights, there are several comments as follows. First, as seen in Figure~\ref{22.1b}, the partial ordering defined on the HaPPY tensor network naturally gives rise to the concept of ``local flow lines’’ or ``arrows”, which strongly suggests that one can glue segments of local flow lines into global entanglement threads. However, at this stage, ~\cite{Pastawski:2015qua} does not specify the concrete rules for such gluing. From the diagram, it seems that there is significant freedom in gluing the local arrows into global entanglement threads. In the quantum circuit representation in Figure~\ref{22.1d}, this degree of freedom is reflected in the fact that the number of wires are conserved in the quantum circuit. Since one can always insert auxiliary two-qudit SWAP gates to redefine the ``apparent wires” ~\footnote{By ``apparent wires'', we refer to the visually parallel straight lines in the standard quantum circuit diagrams, representing the evolution of individual qudits.}, the diagrammatic representation of a quantum circuit is not unique. In Section~\ref{sec42}, we will propose an interesting way to reconcile the non-uniqueness of such circuit drawings with the definite trajectories of entanglement threads, by comparing the concept of entanglement threads with the earlier concept of bit threads~\cite{Freedman:2016zud, Cui:2018dyq, Headrick:2017ucz}.

The issue of how to glue local ``arrows'' into global entanglement threads is in fact equivalent to the question of determining the trajectories of entanglement threads through the bulk. This is the central topic we will address in detail in the following sections. What is important for now is that, in the context of tensor networks, we have already obtained a concrete way to define the trajectories of entanglement threads: the trajectory of a thread can be defined as the ordered sequence of basic tensors (or graphically, ``hexagonal tiles’’) in the bulk that it passes through. Moreover, since each leg extending from a basic tile represents a qudit, each entanglement thread can actually be considered as a collection of qudits. We shall adopt and further develop these insights in the following sections.


\section{Trajectories of holographic entanglement threads in the Bulk}\label{sec2}
\subsection{The ``No-Return'' Rule of Entanglement Threads on RT Surfaces}\label{sec21}

\begin{figure}
     \centering
       \begin{subfigure}[b]{0.46\textwidth}
         \centering
         \includegraphics[width=\textwidth]{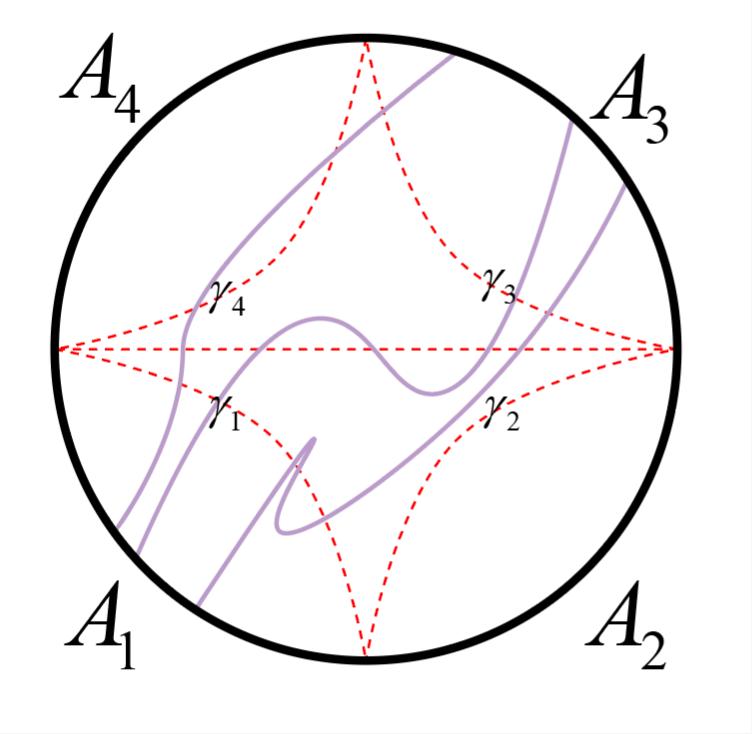}
         \caption{}
         \label{31.1a}
     \end{subfigure}
     \hfill
     \begin{subfigure}[b]{0.45\textwidth}
         \centering
         \includegraphics[width=\textwidth]{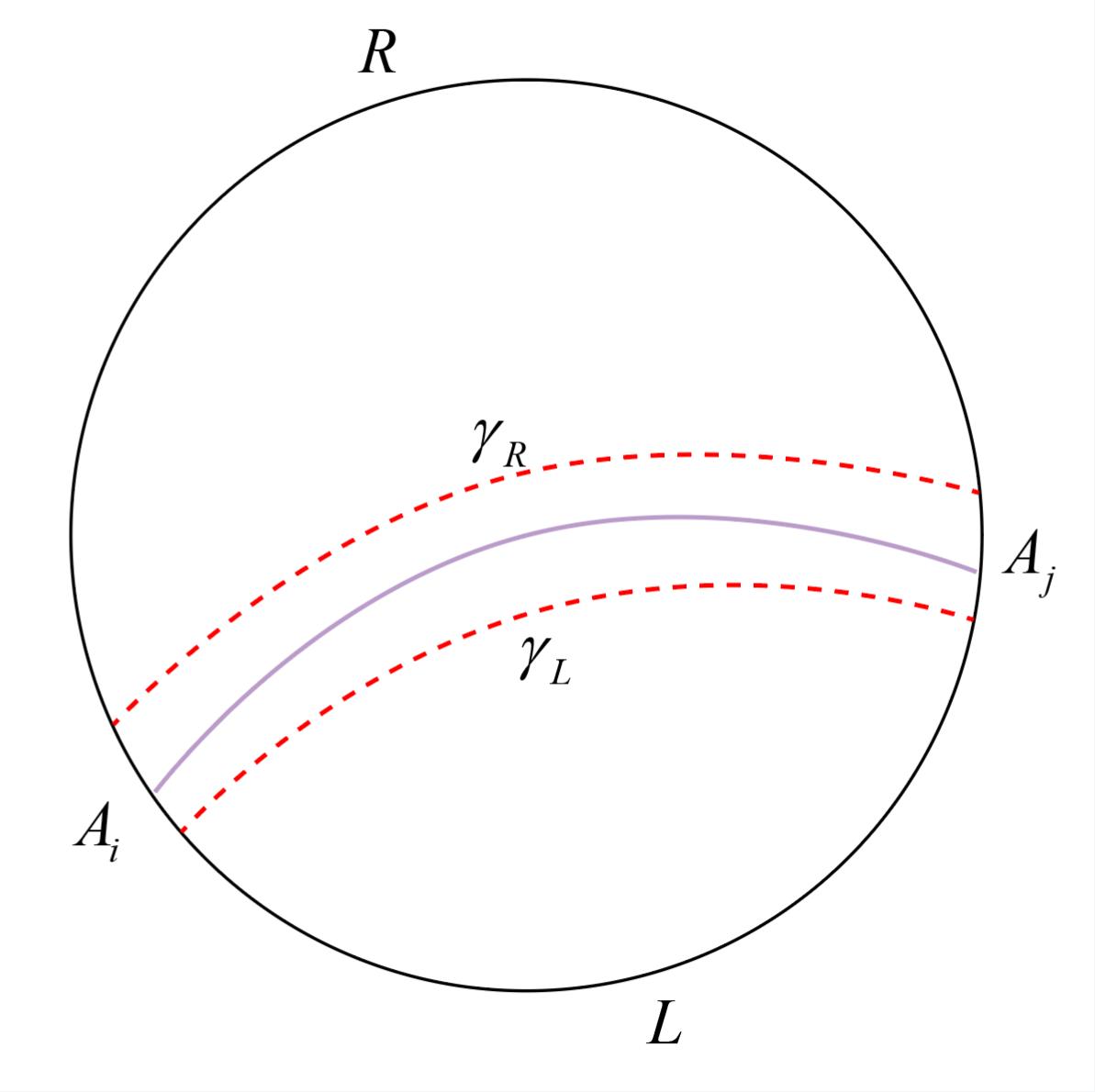}
         \caption{}
         \label{31.1b}
     \end{subfigure}
             \caption{(a) Some counterexamples that violate the thread trajectory rules. (b) By continuously shrinking the sizes of $A_i$ and $A_j$, in the limit, $\zeta_{ij}$ must coincide with the geodesics.}
        \label{31.1}
\end{figure}

As mentioned in the previous section, according to our understanding, each entanglement thread should be constructed by gluing together a serious of adjacent arrows. However, at least in the case of HaPPY tensor network, the partial ordering information characterized by the collection of all local directed arrows is not sufficient to fully determine the specific trajectory of each entanglement thread. The question of how local ``arrows'' are glued into global entanglement threads is essentially equivalent to the question of how entanglement threads traverse the bulk.

To come straight to the point, in this section, we will further demonstrate that the trajectories of these entanglement threads are precisely geodesics in the holographic bulk. This conclusion only requires us to pay attention to the following phenomenological fact: according to the thread scheme described in Section~\ref{sec11}, an entanglement thread cannot pass through a given simply connected RT surface more than once. Figure~\ref{31.1a} illustrates some examples of disallowed entanglement thread configurations. The reason lies in the constraint from the holographic principle. For convenience, we adopt the unit convention $ 4G_N = 1 $. Under this convention, the entanglement entropy between a simply connected region $ A $ and its complement $ \bar{A} $ is exactly equal to the area of the RT surface $ \gamma_A $. Since we require that the number of entanglement threads connecting $ A $ and $ \bar{A} $ be precisely equal to this entropy, it follows that each unit area on the RT surface $ \gamma_A $ can accommodate exactly one entanglement thread, and each thread is allowed to occupy that unit area only once. Similarly, an entanglement thread with both endpoints in $ A $, or both in $ \bar{A} $, is not allowed to pass through $ \gamma_A $, because the area of $ \gamma_A $ is only sufficient to accommodate ``travelers'' whose starting and ending points lie in $ A $ and $ \bar{A} $, respectively.

According to this ``no-return'' rule of entanglement threads on RT surfaces, it is not hard to conjecture that the trajectories of entanglement threads in the bulk should coincide with geodesics. To intuitively understand this conjecture requires only a few lines of argument. As shown in Figure~\ref{31.1b}, let us suppose that the sizes of two elementary regions $ A_i $ and $ A_j $ are made very small. Denote the regions between them as $ L $ and $ R $, respectively. Then, according to the no-return rule, the entanglement thread $ \zeta_{ij} $ connecting $ A_i $ and $ A_j $ will be confined within a narrow channel bounded by two RT surfaces $ \gamma_L $ and $ \gamma_R $, because $ \zeta_{ij} $ is not allowed to pass through either $ \gamma_L $ or $ \gamma_R $ more than once. Therefore, as we continuously shrink the sizes of $ A_i $ and $ A_j $, in the limiting case, the trajectory $ \zeta_{ij} $ must coincide with the geodesic segment defined between $ \gamma_L $ and $ \gamma_R $.

\subsection{The Entanglement Meaning of Thread Trajectories}\label{sec22}

\begin{figure}
    \centering
    \includegraphics[scale=0.15]{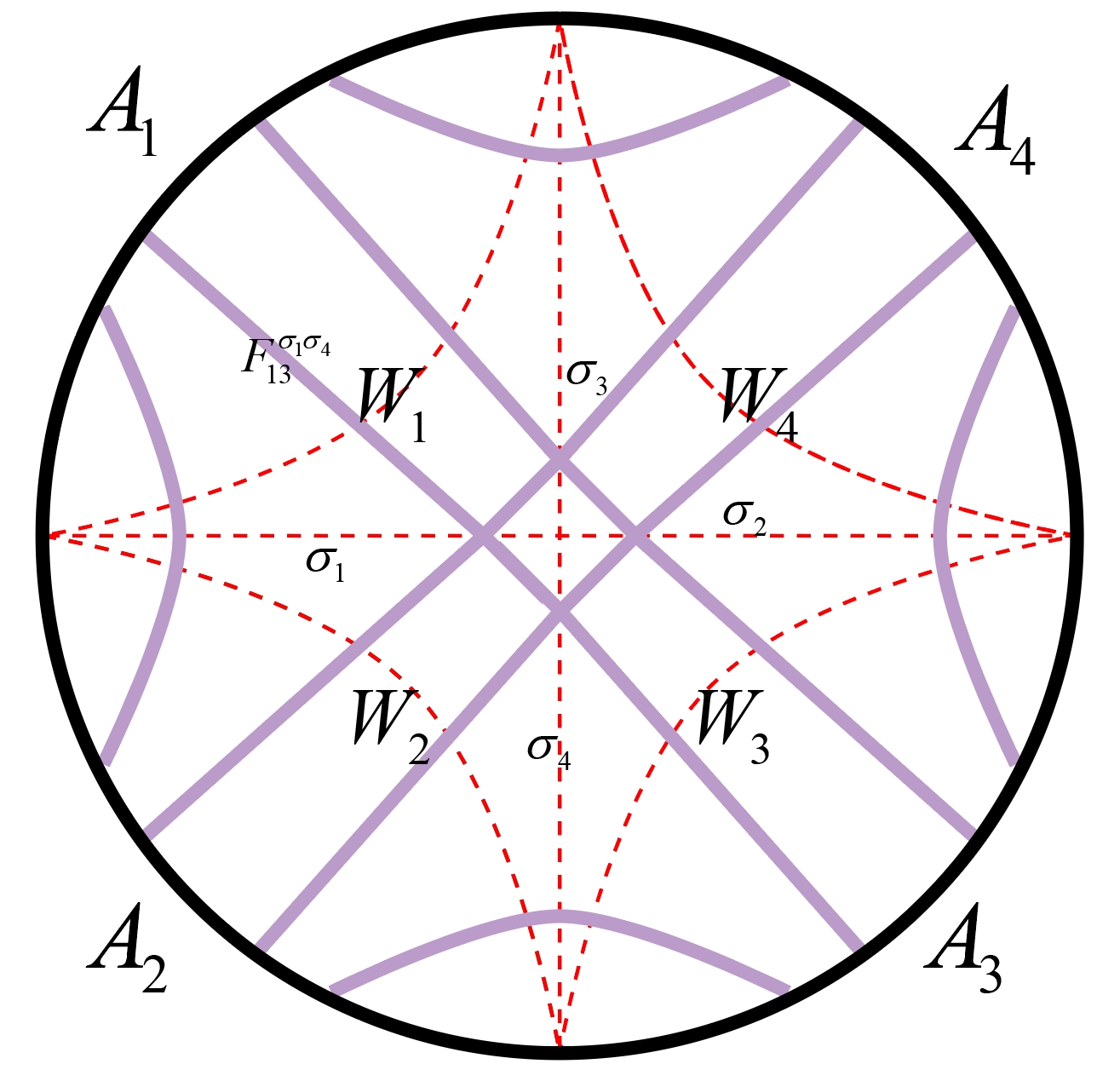}
    \caption{A decomposition of the bulk slice $ W $ into four parts $ W_1, W_2, W_3, W_4 $ by two intersecting RT surfaces $ {\gamma _{12}} = {\sigma _3}  \cup  {\sigma _4} $ and $ {\gamma _{23}} = {\sigma _1} \cup {\sigma _2} $. Further considering the trajectory of the entanglement threads in each fragment, now a total of 8 independent thread bundles (each bundle is represented by a thick purple line) are required, and they are marked using the generalized RT surfaces (bulk extremal surfaces) they pass through.}\label{32.1}
\end{figure}

Although we have already argued from a phenomenological perspective that entanglement threads follow certain trajectories, it is still necessary to discuss their precise meaning at the level of quantum information theory, so as to provide a more convincing foundation for this concept. In this subsection, with the intuition of tensor network picture in mind, we present an analysis from the quantum state perspective concerning the trajectories of entanglement threads in the bulk. Note that our following discussion does not rely on assuming the AdS bulk to be exactly modeled by a HaPPY tensor network; it can be fully applied to more abstract and general classes of tensor networks. This is one of the advantages of the entanglement thread approach over specific tensor network constructions.

The basic idea is to locally decompose the entire bulk. Without loss of generality, we consider (as shown in Figure~\ref{32.1}) a decomposition of the bulk slice $ W $ into four parts $ W_1, W_2, W_3, W_4 $ by two intersecting RT surfaces $ \gamma_{12} $ and $ \gamma_{23} $. This can also be viewed as a four-part division of a holographic tensor network, as long as we conventionally define RT surfaces as minimal cuts of internal bonds within the tensor network. Let us, for instance, focus on the small region $ W_1 $ adjacent to $ A_1 $, whose boundary is labeled as $ \partial W_1 = A_1 \cup \sigma_1 \cup \sigma_3 $. Treating the bulk slice as a tensor network~\footnote{Regardless of whether the HaPPY network is a precise model, one can always imagine a sufficiently good holographic tensor network to make the discussion satisfactory. We note that recent work~\cite{Hung:2024gma, Geng:2025efs, Chen:2022wvy} has made a remarkable step toward constructing such concrete tensor networks.}, we are led to two insights. First, the boundary $ A_1 \cup \sigma_1 \cup \sigma_3 $ can be regarded as corresponding to a pure state, which in principle can be considered as constructed from contracting the basic tensors within the subnetwork $ W_1 $. This idea is referred to as the surface-state correspondence in~\cite{Miyaji:2015fia, Miyaji:2015yva}. Second, one can imagine that the local arrows in $ W_1 $ are sufficient to be glued into a local entanglement thread configuration $ V_{W_1} $, which characterizes the entanglement structure of the sub-bulk region $ W_1 $ and its boundary dual state—just as when we consider the full bulk region $ W $, all the local arrows can be glued into a global thread configuration $ V_W $. In particular, $ V_{W_1} $ should be a local portion of the full thread configuration $ V_W $, since they share the same set of local arrow “material” in region $ W_1 $.

After abstracting away from specific tensor network models and returning to the pure AdS setting, the above argument motivates the following proposal: as illustrated in Figure~\ref{32.1}, when the entire holographic boundary is partitioned into four regions~\footnote{Of course, in our framework, one can expect more refined thread configurations when the boundary is partitioned into more regions.}, we should at least be able to construct a compatible thread configuration, such that: not only should it reproduce, via the prescription in Section~\ref{sec11}, the holographic entanglement entropies for regions $ A_1, A_2, A_3, A_4 $ and their unions related to the full bulk slice $ W $, but also, for each $ W_i $, it should reproduce the holographic entanglement entropies of the related boundary regions (e.g., for $ W_1 $: $ A_1, \sigma_1, \sigma_3 $, and their unions), using the same prescription. 

Now we proceed with the analysis. From a global perspective, following the formula in Section~\ref{sec11}, to characterize the entanglement structure among the four elementary regions $ A_1, A_2, A_3, A_4 $, we only need six independent bundles: $ \{F_{12}, F_{13}, F_{14}, F_{23}, F_{24}, F_{34}\} $, which are sufficient to capture the set of entropies $ \{S_1, S_2, S_3, S_4, S_{12}, S_{23}\} $. Furthermore, in the metaphor of tensor networks discussed in Section~\ref{sec12}, solving for these six values of $ F_{ij} $ amounts to determining the number of entanglement threads connecting each pair $ (A_i, A_j) $, assuming that the local arrows have been glued into a global configuration. However, this does not yet specify the exact paths these threads take through the bulk network. Now, by decomposing $ W $ into four regions $ W_1, W_2, W_3, W_4 $, and applying the entanglement thread formula to each subregion, we gain further information about the trajectories of the threads. For example, consider $ W_1 $ and its three boundary regions $ A_1, \sigma_1, \sigma_3 $. Constructing $ V_{W_1} $ requires, in principle, three thread bundles with fluxes denoted by $ \{F_{A_1\sigma_1}, F_{A_1\sigma_3}, F_{\sigma_1\sigma_3}\} $. However, since $ V_{W_1} $ is just a fragment of $ V_W $, we expect the entanglement entropies of $ A_1, \sigma_1, \sigma_3 $ to be fully captured using only the threads already present in $ V_W $. This in turn constrains the possible trajectories of the threads in $ V_W $. To see this, observe that, by the no-return rule for entanglement threads, a bundle going from $ A_1 $ to $ A_3 $ has two possible routes: via $ \sigma_3 \to \sigma_2 $ or via $ \sigma_1 \to \sigma_4 $. Likewise, the bundle from $ A_2 $ to $ A_4 $ has two possible paths. We can write:

\begin{equation}
\begin{aligned}
F_{13} &= F_{13}^{\sigma_1\sigma_4} + F_{13}^{\sigma_3\sigma_2}, \\
F_{24} &= F_{24}^{\sigma_1\sigma_3} + F_{24}^{\sigma_4\sigma_2}.
\end{aligned} \label{dec}
\end{equation}
Here, the superscripts indicate the intermediate surfaces traversed by the threads. On the other hand, according to the no-return rule, the threads from $ A_1 $ to $ A_2 $ must all pass through $ \sigma_1 $, and cannot pass through $ \sigma_2 $, or else they would cross the RT surface $ \gamma_{12} $ at least twice. Applying similar reasoning, we now have eight independent thread bundles in total, that is :$\{F_{12}^{\sigma_1},\quad F_{23}^{\sigma_4},\quad F_{34}^{\sigma_2},\quad F_{14}^{\sigma_3},\quad F_{13}^{\sigma_1\sigma_4},\quad F_{13}^{\sigma_3\sigma_2},\quad F_{24}^{\sigma_1\sigma_3},\quad F_{24}^{\sigma_4\sigma_2}.\}$. We can now solve for the flux values of these eight bundles. This is because, in addition to the six entanglement entropies of the boundary regions of $ W $ (which correspond to the areas of RT surfaces $ \gamma_1, \gamma_2, \gamma_3, \gamma_4, \gamma_{12}, \gamma_{23} $), we now also have entanglement entropies of the boundary subregions of $ W_1, W_2, W_3, W_4 $ as further constraints. For example, for $ W_1 $, the entanglement entropy of $ \sigma_1 $ (whose RT surface is itself, hence the entropy equals its area) provides a constraint:
\begin{equation}
\text{Area}(\sigma_1) = F_{12}^{\sigma_1} + F_{13}^{\sigma_1\sigma_4} + F_{24}^{\sigma_1\sigma_3}.
\end{equation}
Note that since the areas of all six simply connected RT surfaces have already been used, once the area of $\sigma_1$ is determined, the area of $\sigma_2$ is also determined, since their sum is exactly the area of $\sigma_23$. We are then left with only one last constraint, which can be chose as, for instance, the entropy of $ \sigma_3 $ (as a boundary subregion of $ W_4 $):
\begin{equation}
\text{Area}(\sigma_3) = F_{14}^{\sigma_3} + F_{13}^{\sigma_3\sigma_2} + F_{24}^{\sigma_1\sigma_3}.
\end{equation}
Finally, we obtain eight effective constraints to solve for the eight flux values.~\footnote{Here we omit the other six equations corresponding to the RT surfaces $ \gamma_1, \gamma_2, \gamma_3, \gamma_4, \gamma_{12}, \gamma_{23} $, since they are already self-evident according to Section~\ref{sec11}.} In particular, we can calculate the flux values of the splitting flow from ${F_13}$ and ${F_24}$ explicitly as:
\begin{equation}
\begin{aligned}
F_{13}^{\sigma_1\sigma_4} &= \frac{1}{2} \left( \text{Area}(\sigma_1) + \text{Area}(\sigma_4) - \text{Area}(\gamma_2) \right), \\
F_{13}^{\sigma_3\sigma_2} &= \frac{1}{2} \left( \text{Area}(\sigma_2) + \text{Area}(\sigma_3) - \text{Area}(\gamma_4) \right), \\
F_{24}^{\sigma_1\sigma_3} &= \frac{1}{2} \left( \text{Area}(\sigma_1) + \text{Area}(\sigma_3) - \text{Area}(\gamma_1) \right), \\
F_{24}^{\sigma_4\sigma_2} &= \frac{1}{2} \left( \text{Area}(\sigma_2) + \text{Area}(\sigma_4) - \text{Area}(\gamma_3) \right).
\end{aligned} \label{mutu}
\end{equation}
We find that each formula in \eqref{mutu} corresponds exactly to the conditional mutual information between two boundary regions of a sub-bulk. For instance, $ F_{24}^{\sigma_1\sigma_3} $ is precisely the conditional mutual information between $ \sigma_1 $ and $ \sigma_3 $ within $ W_1 $. This shows that, within $ W_1 $, the entanglement threads connecting $ \sigma_1 $ and $ \sigma_3 $ are entirely derived from the portion of the global thread bundle connecting $ A_2 $ and $ A_4 $ that passes through those surfaces—just as we would expect.

We have come full circle to present the following argument to the reader: Not only can we, in the sense of Section~\ref{sec11}, discuss the flux values of entanglement threads between elementary boundary regions (which rely only on the topological structure of threads, and can be simply encoded in a complete graph), but we can also discuss how threads traverse the bulk slice—and show that their precise trajectories can, in principle, be determined from more detailed entanglement information within the bulk. In Section~\ref{sec21}, we boldly conjectured that the trajectories of entanglement threads are geodesics. In Section~\ref{sec3}, we will use the language of kinematic space~\cite{Czech:2015kbp,Czech:2015qta} to verify that assuming threads follow geodesics indeed leads to the result in \eqref{mutu}.

\subsection{Trajectories of Entanglement Threads in the HaPPY Tensor Network}\label{sec23}

As an example of the close connection between the entanglement thread approach and the tensor network approach, let us now, based on the analysis of Section~\ref{sec22}, proceed to rigorously determine the trajectories of entanglement threads in the HaPPY tensor network model, extending the results of \cite{Pastawski:2015qua}. The insight from Section~\ref{sec22} is to consider a subregion of the bulk—or in the tensor network context, a subnetwork. Now let us take the subnetwork to be a single six-legged tensor $ T $ itself. Applying the labeling and orientation rule from Section~\ref{sec12}—where numbers are assigned to the dual graph—we find that the six legs of $ T $ must be oriented such that its three incoming legs are adjacent to each other. Accordingly, we can assign the six indices $ a, b, c, d, e, f $ (physically representing six qudits) to the six legs in a unique way, starting from the first incoming leg and proceeding counterclockwise. Again, we observe that, in principle, there is more than one way to connect arrows of the same direction into thread-like objects. However, if we treat the six qudits as six elementary regions and consider all simply connected regions such as $ a \cup f $, $ a \cup f \cup e $, etc., then—similar to the argument in Section~\ref{sec22}—we can compute that, within this sub-tensor network, we have:
\begin{equation}
F_{ad} = F_{be} = F_{cf} = \log d, \label{logd}
\end{equation}
where $ d $ is the dimension of each qudit, and all other fluxes are zero. Let $ d = 2^{\chi} $, then Equation~\eqref{logd} implies that, in the HaPPY tensor network, for each subnetwork represented by a single tensor, the entanglement thread distribution is that: there are $\chi$ threads connecting $a$ to $d$, $\chi$ threads connecting $b$ to $e$ and $\chi$ threads connecting $c$ to $f$ (see Figure~\ref{22.1c}. Only in this way can the flux of threads connecting each simply connected region and its complement correctly match the entanglement entropy. For the HaPPY tensor network, this simple analysis fully determines the concrete configuration of entanglement threads, as shown in Figure~\ref{22.1a}. Note that since this tensor network is merely a discrete depiction of a spatial slice of pure AdS, its entanglement thread configuration inevitably loses some finer details. Nevertheless, we observe that each entanglement thread in the figure indeed appears as a geodesic, consistent with our argument in Section~\ref{sec21}.

Recall that, as reviewed in Section~\ref{sec12}, the HaPPY tensor network can be topologically deformed into a conventional quantum circuit diagram. We can now also mark the entanglement thread trajectories in the quantum circuit diagram. Figure~\ref{22.1d} shows examples of a few entanglement threads and their corresponding paths. It is evident that, since we arbitrarily chose a particular representation of the apparent wire set in the quantum circuit, these entanglement thread trajectories are not simple straight lines. In principle, one can choose a canonical representation of the quantum circuit such that each apparent wire precisely corresponds to a single entanglement thread. In the next section, we will use tools from kinematic space to explicitly construct such a canonical representation.


\section{Entanglement Threads, Kinematic Space, and Quantum Circuits}\label{sec3}
\subsection{Characterizing Entanglement Threads through Kinematic Space}\label{sec31}

We have proposed in Section~\ref{sec2} that the traveling trajectory of an entanglement thread in the bulk is exactly a geodesic. This naturally relates to the concept of kinematic space~\cite{Czech:2015kbp,Czech:2015qta}, which is defined as the set of all boundary-anchored geodesics in the holographic bulk! Indeed, we should have expected this, because in the formulation of kinematic space, conditional mutual information plays a central role in the metric of the space due to its elegant mathematical structure. However, we arrived here from a different direction—namely, from the perspective of quantum entanglement. We place a review of the kinematic space in the appendix \ref{appa}. Here, to verify that the trajectories of entanglement threads are indeed geodesics, we can adopt the mathematical techniques used to characterize geodesics in kinematic space to compute the fluxes in the example of Figure \ref{32.1}.

\begin{figure}
     \centering
     \begin{subfigure}[b]{0.46\textwidth}
         \centering
         \includegraphics[width=\textwidth]{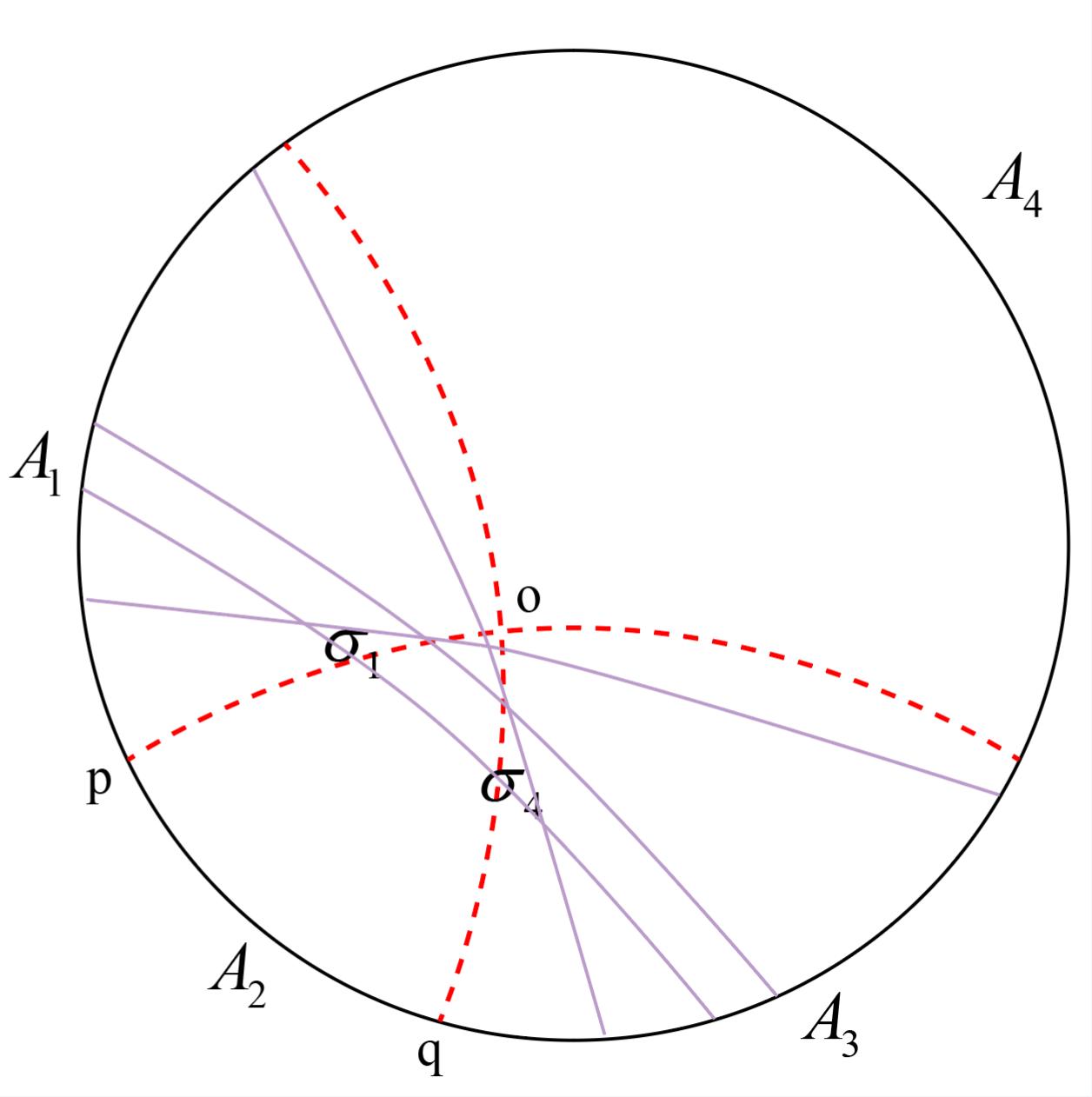}
         \caption{}
         \label{41.1a}
     \end{subfigure}
     \hfill
     \begin{subfigure}[b]{0.46\textwidth}
         \centering
         \includegraphics[width=\textwidth]{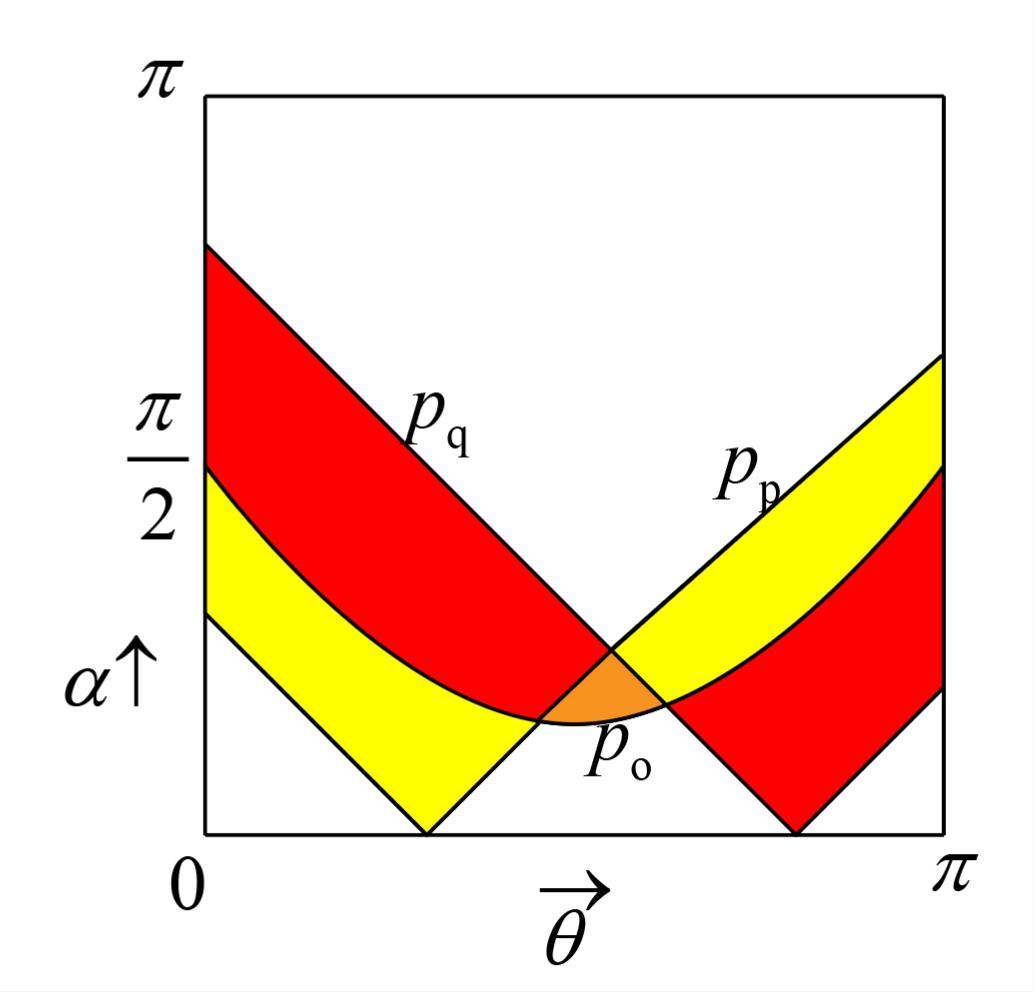}
         \caption{}
         \label{41.1b}
     \end{subfigure}
     \caption{(a)	Some geodesics accounting for $F_{13}^{\sigma_1 \sigma_4}$ are shown. (b) The set of geodesics contributing to $ F_{13}^{\sigma_1 \sigma_4} $ corresponds to the orange region, which is the intersection of the red and yellow regions.}
     \label{41.1}
\end{figure}

As a demonstration, let us compute the flux $F_{13}^{\sigma_1 \sigma_4}$, which denotes the number of threads starting from $ A_1 $, passing through the surfaces $ \sigma_1 $ and $ \sigma_4 $, and reaching $ A_3 $. Since we now assume that the trajectories of these entanglement threads precisely coincide with geodesics, it becomes relatively straightforward to identify the set of geodesics just enough to satisfy the trajectory constraints, which, in kinematic space, form particular curves made up of collections of points. These borderline curves in kinematic space correspond precisely to the so-called point-curves associated with the three points involved in the problem, namely $ o $, $ p $, and $ q $, as illustrated in Figure~\ref{41.1}. In Figure~\ref{41.1}b, the red region is bounded by the point-curves corresponding to $ o $ and $ q $, representing those geodesics in the original space that intersect the segment $ oq $. The yellow region is bounded by the point-curves corresponding to $ o $ and $ p $, representing geodesics intersecting the segment $ op $ in the original AdS space. Clearly, the geodesics that contribute to $ F_{13}^{\sigma_1 \sigma_4} $ are those lying in the intersection of the red and yellow regions—that is, the orange region shown in the figure. The area of the orange region directly gives the value of $ F_{13}^{\sigma_1 \sigma_4} $. Using the notation from Appendix \ref{appa}, we can compute:
\begin{equation}\label{f13k}
F_{13}^{\sigma_1 \sigma_4} = \frac{1}{2} \left( \frac{1}{4} \int\limits_{p_o \Delta p_q} \tilde{\omega} + \frac{1}{4} \int\limits_{p_o \Delta p_p} \tilde{\omega} - \frac{1}{4} \int\limits_{p_p \Delta p_q} \tilde{\omega} \right). \end{equation}
From equation (\ref{suv}), we can identify that the last term inside the parentheses is precisely $S_2 = \frac{\text{Area}(\gamma_2)}{4 G_N}$, and we can also recognize that the first two terms are exactly the same as the corresponding terms in equation (\ref{mutu}). Thus, (\ref{f13k}) is completely consistent with (\ref{mutu}).

We have therefore verified that the mathematics of kinematic space can indeed be correctly applied to study the fluxes of entanglement threads. This provides us with a systematic mathematical tool for analyzing the entanglement structure in holographic duality using holographic entanglement threads. The simple example shown here is not sufficient to fully demonstrate the power of the mathematical structure of kinematic space. However, for more general problems—such as more complex divisions of boundary regions, or more general holographic scenarios like wormholes—the simple method of using linear combinations of the areas of a few RT surfaces in the bulk will no longer suffice. In such cases, the mathematical language of integral geometry will offer powerful tools. We leave these computational challenges for future exploration.

\subsection{Kinematic Space, Quantum Circuits, and Complexity}\label{sec32}

\begin{figure}
    \centering
    \includegraphics[scale=0.2]{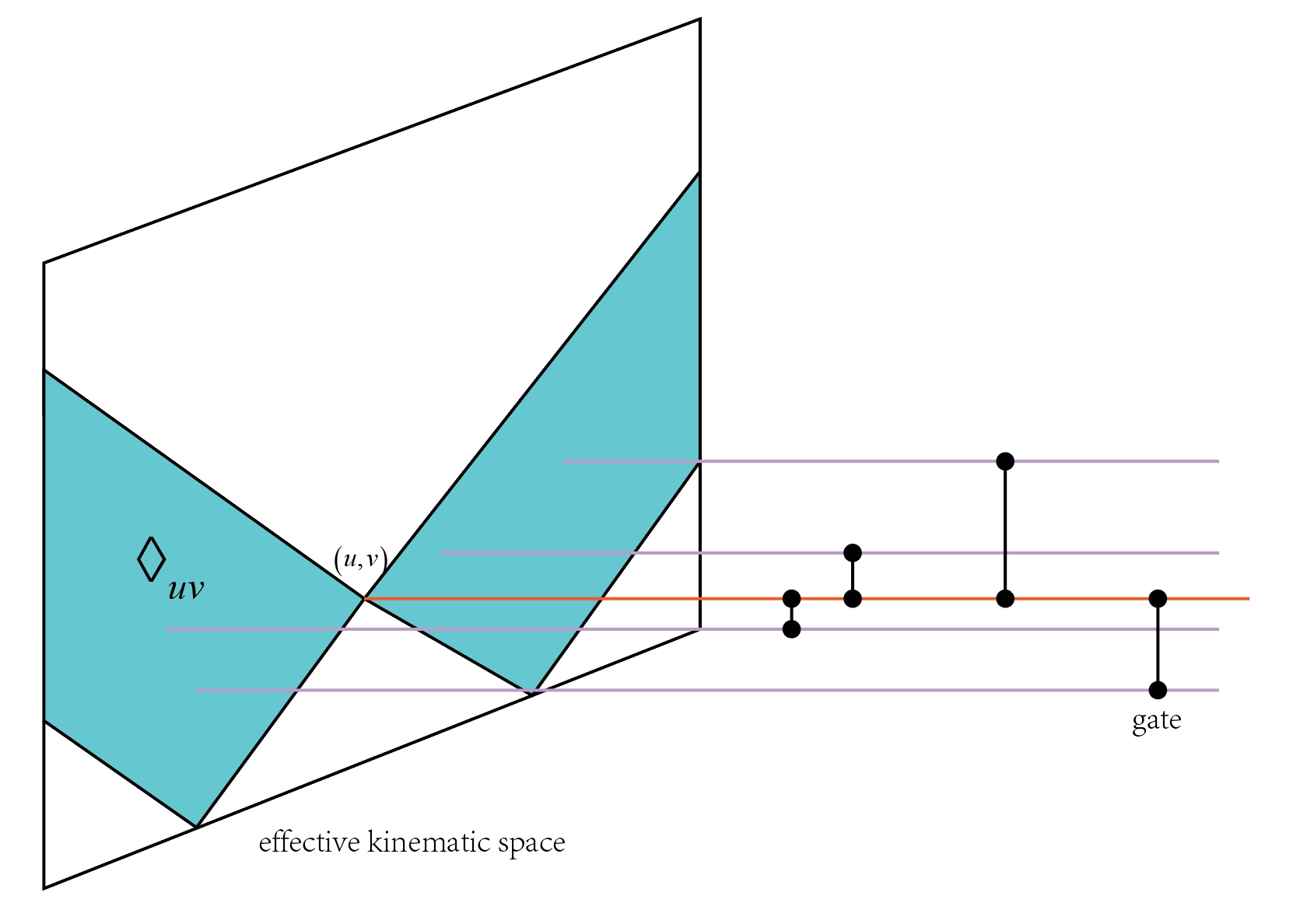}
    \caption{Regarding kinematic space as an input board, a qubit $(u,v)$ will be coupled through a quantum gate with each wire emanating from the $\diamondsuit_{uv}$ region.}
    \label{42.1}
\end{figure}

Identifying each entanglement thread's trajectory as a geodesic in the bulk is a non-trivial assertion. Although we have seen various roles that geodesics play in different ``thread-like" pictures in the literature—such as bit threads, string condensations~\cite{Yan:2019vzp}, and PEE threads~\cite{Lin:2024dho,Lin:2023rxc}—in our framework, this conclusion is reached based on an assumption that carries a direct and explicit meaning in quantum information theory, rather than an indirect one. In the following, we will demonstrate that, using the language of kinematic space, we can construct a canonical quantum circuit in which each apparent wire precisely corresponds to an entanglement thread in the bulk—that is, a geodesic. In other words, each geodesic in the original tensor network is ``straightened” into a wire in this canonical quantum circuit, while each tensor is bisected and mapped to a quantum gate. Specifically, when two geodesics intersect in the original tensor network, they are entangled through a basic tensor (or a simple combination of a few basic tensors).~\footnote{ Taking the HaPPY tensor network as an example, as shown in Figure~\ref{22.1}, two intersecting entanglement thread bundles are coupled within a single 3-to-3 quantum gate. However, it can be shown that such a 3-to-3 perfect tensor gate can be decomposed into a composition of three two-qudit quantum gates~\cite{Pozsgay:2023oyc} (see Figure~\ref{52.1}).} Conversely, in the resulting quantum circuit, whenever two wires are coupled through a quantum gate (or a combination of a few basic quantum gates), it means that the corresponding geodesics in the original tensor network intersect.

Now we turn to figure~\ref{42.1}. Since each point in kinematic space $K$ corresponds precisely to a geodesic, kinematic space provides a convenient way to arrange the relative positions of wires after the topological deformation of the tensor network into a quantum circuit. In other words, kinematic space can be viewed as a ``input board" for the quantum circuit, where each point $(u, v)$ corresponds to one of the input qudits.~\footnote{ We can use a discrete picture to describe this precisely: first, regularize kinematic space by dividing it into small diamonds with volume equal to 1. Thus, each small diamond corresponds to one thread. On the other hand, recall that the RT formula applies to regions far larger than the Planck scale $\ell_p$, therefore each diamond $[\diamondsuit_{A_i, A_j | L}]$, corresponding to regions far larger than $\ell_p$, can be viewed as a fundamental cell that contains considerable number of threads therein.} These input qudits extend into wires, with different wires entangled through quantum gates. Compared to Figure \ref{22.1d}, which is a relatively arbitrary deformation of the original tensor network, it can be seen that the wires in the current quantum circuit are perfectly arranged as a series of parallel straight lines, conforming to the usual convention in quantum circuit literature. This indicates that the kinematic space is a very elegant way of organizing the entanglement structure. The special arrangement of qubits on the kinematic space has the advantage of expediently describing how the wires are coupled through quantum gates. As shown in figure~\ref{42.1}, consider the qubit $(u,v)$, which in the original space $N$ represents a geodesic $\varsigma_{uv}$ connecting two anchor points $u$ and $v$ on the boundary. Which wires will couple with this qubit? We have already argued that it should be the set of other geodesics that intersect $\varsigma_{uv}$, which corresponds precisely to the diamond-shaped region $\diamondsuit_{uv}$ in kinematic space $K$. In a word, regarding kinematic space as an input board, a qubit $(u,v)$ will be coupled through a quantum gate with each wire emanating from the $\diamondsuit_{uv}$ region. It turns out that this conclusion leads us to the concept of holographic complexity.

Let us briefly review the concept of complexity. As the name suggests, in the context of quantum circuits, the complexity of a state is the minimum number of gates required to transform a reference state into the target state. The complexity of an operator is the minimum number of gates required to simulate that operator using a quantum circuit. For quantum theories defined in finite-dimensional Hilbert spaces, this definition is straightforward. However, defining the complexity of a field theory state is notoriously challenging. Nevertheless, in the context of holographic duality, the complexity of a field theory has been proposed to have a bulk dual. In the first wave of proposals for holographic duals of field theory complexity,\cite{Susskind:2014rva, Stanford:2014jda, Brown:2015bva, Brown:2015lvg} considered a system compose of two copies of CFT initially entangled through a thermofield double state. Then the complexity of the system evolving in time is argued to be holographic dual to either the volume of the Einstein-Rosen bridge of the dual two-sided black hole or the action of a Wheeler-DeWitt patch. The first proposal, which relates complexity to volume, is known as the ``complexity = volume" (CV) proposal~\cite{Susskind:2014rva, Stanford:2014jda}. Based on complexity = volume, \cite{Carmi:2016wjl, Alishahiha:2015rta}~also proposed the subregion complexity, which states that for a boundary subregion $A$, the volume of the bulk region enclosed by its corresponding RT surface gives the complexity of the reduced density matrix for $A$. Specifically, when $A$ is taken to be the entire boundary system, the original CV proposal is recovered.

\begin{figure}
    \centering
    \includegraphics[scale=0.26]{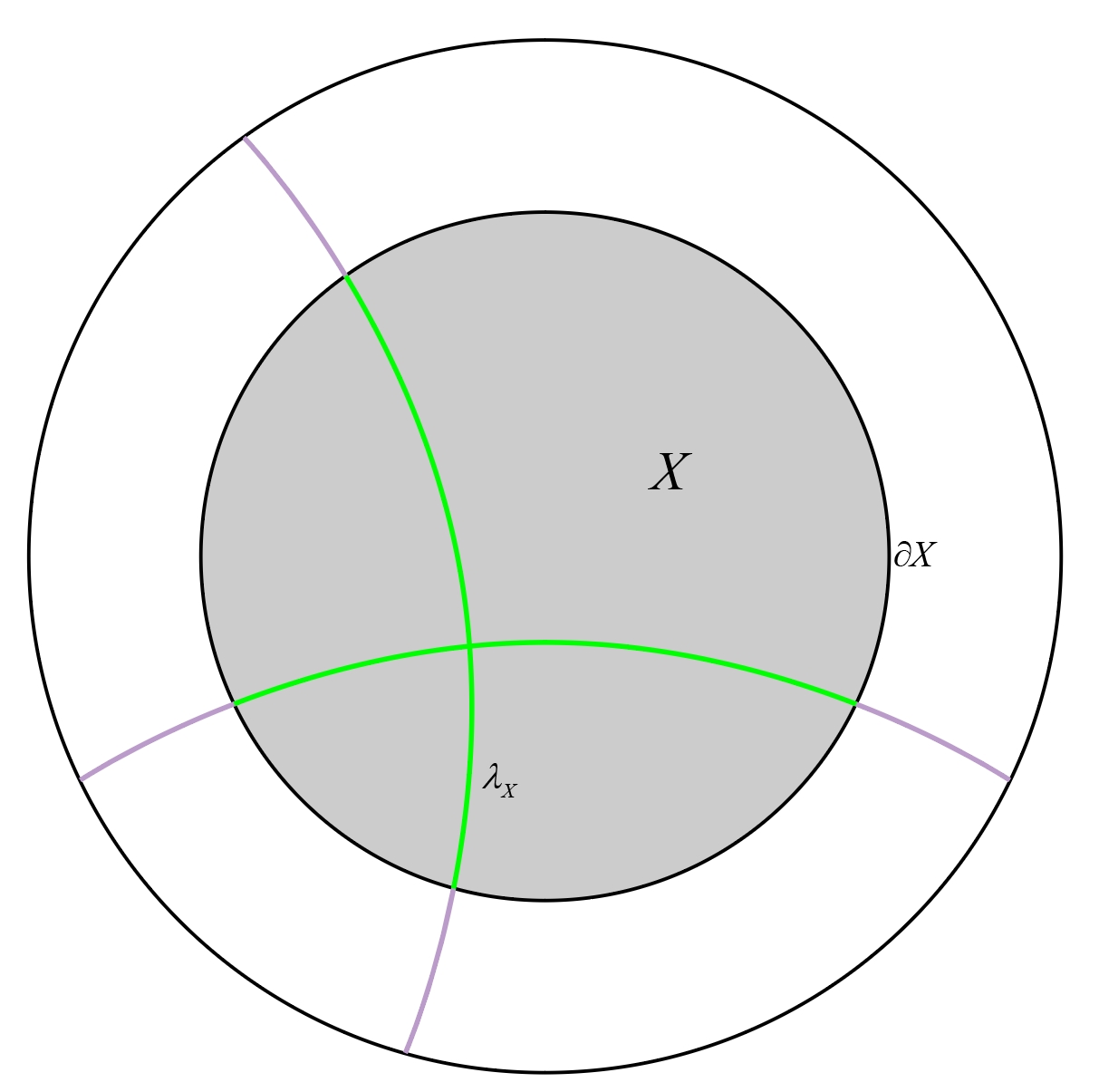}
    \caption{The volume of a bulk region $X$ (marked in grey) can be given by an integral over the length of all geodesic chords (marked in green). }
    \label{42.2}
\end{figure}

Returning to our context, a useful formula is the expression of CV complexity in the language of kinematic space~\cite{Abt:2017pmf, Abt:2018ywl}. As shown in figure~\ref{42.2}, the formula states that the volume of a bulk region $X$ can be given by an integral over the length of all geodesic chords. Here, the chord length of a geodesic refers to the length of the segment of the geodesic that lies within $X$ (the green line in the figure). More explicitly:
\be\label{vol}
\frac{{\text{vol}(X)}}{{4 G_N}} = \frac{1}{{2\pi}}\int_{{G_X}} \lambda_X \omega,\ee
where ${G_X}$ is the set of all geodesics $\zeta \in K$ that intersect ${X}$, and $\lambda_X$ is the length of the chord $\zeta \cap X$ ~\footnote{ Equation (\ref{vol}) is a standard result in integral geometry, see Chapter 17 of~\cite{integral} for example.}. Applying (\ref{vol}) to the entire holographic bulk region $X$ gives an expression for global CV complexity, while applying it to a subregion's entanglement wedge $W(A)$ yields an expression for subregion CV complexity.

The quantum circuit interpretation of kinematic space explicitly explain the meaning of CV complexity as expressed in (\ref{vol}). Let us begin by considering the case where $X$ is taken as the entire holographic bulk region. In this case, (\ref{vol}) indicates that CV complexity should be given by the total length of all geodesics in the bulk. In other words, $G_X$ is the set of all geodesics, and $\lambda_X$ is the full length of each geodesic. In our explanation in figure~\ref{42.1}, the right-hand side of (\ref{vol}) essentially counts the number of quantum gates in the quantum circuit deformed from the tensor network. Because for each geodesic $(u,v)$ acting as a wire, the number of quantum gates it couples with is exactly given by the number of wires in the region $\diamondsuit_{uv}$, which also corresponds to the length of the geodesic $\varsigma_{uv}$. As we sum over each wire, the total number of quantum gates in the circuit naturally emerges as the sum of the gates that each wire couples with. Thus (\ref{vol}) explains the complexity.

\begin{figure}
     \centering
       \begin{subfigure}[b]{0.34\textwidth}
         \centering
         \includegraphics[width=\textwidth]{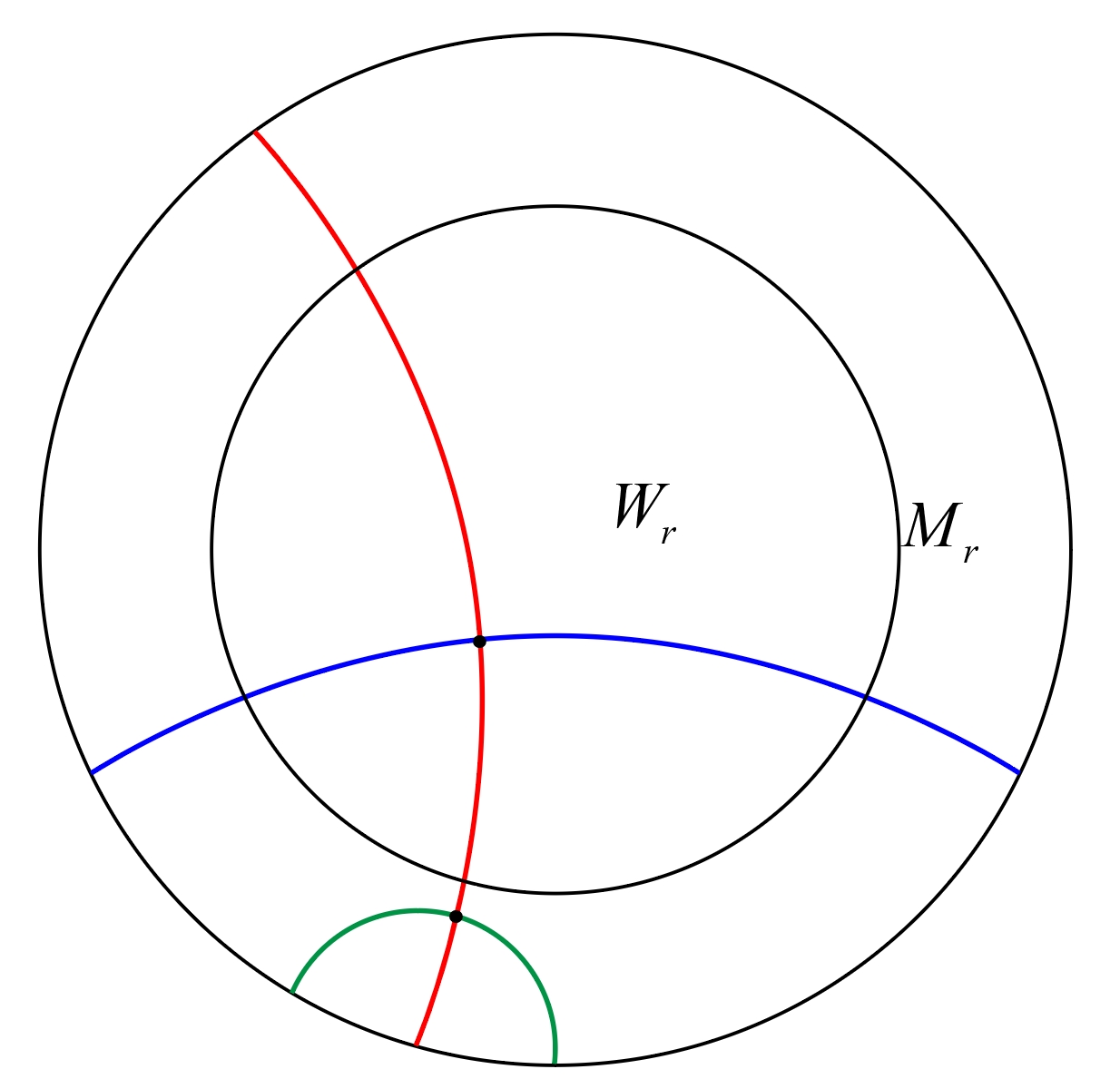}
         \caption{}
         \label{42.3a}
     \end{subfigure}
     \hfill
     \begin{subfigure}[b]{0.44\textwidth}
         \centering
         \includegraphics[width=\textwidth]{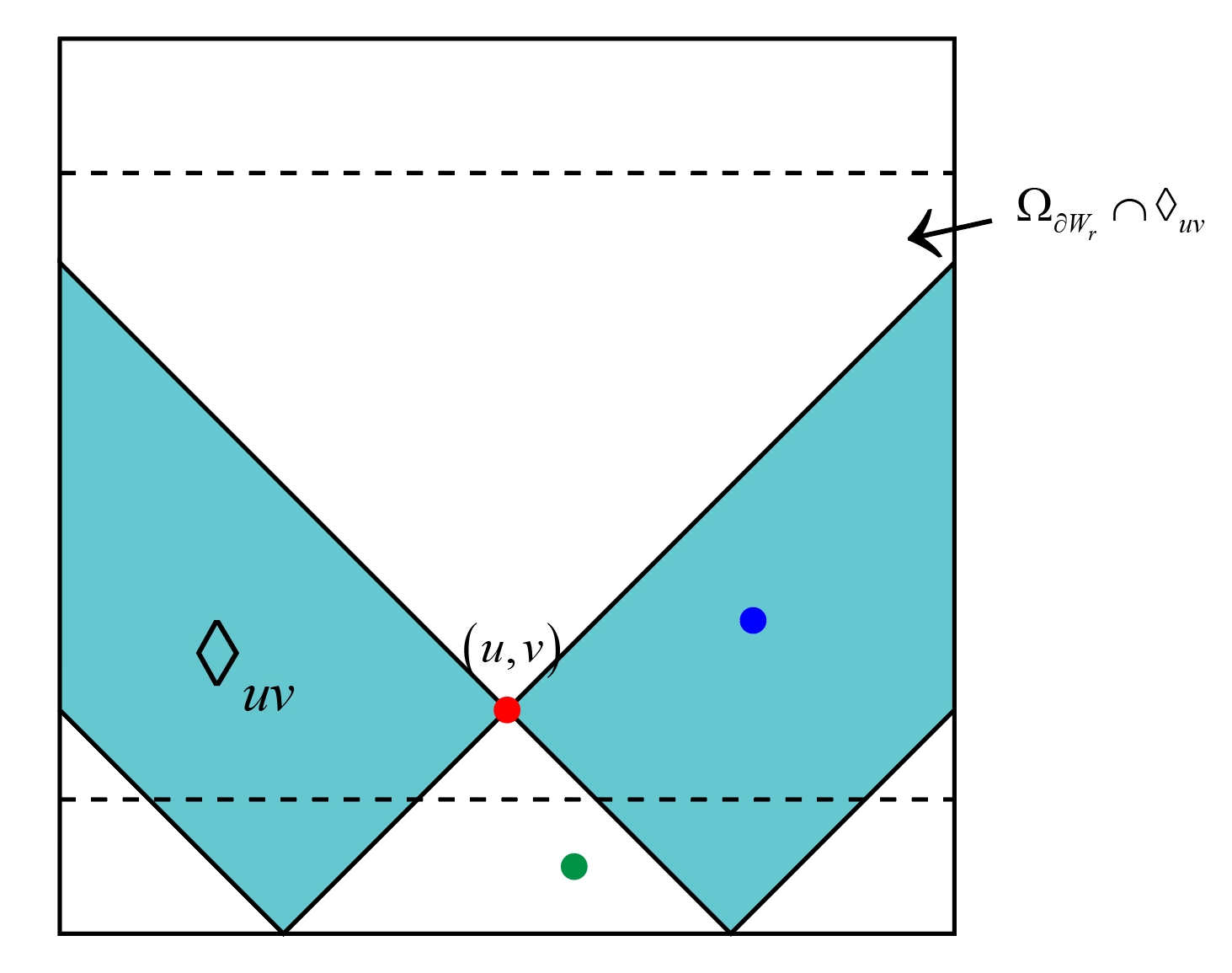}
         \caption{}
         \label{42.3b}
     \end{subfigure}
             \caption{(a) Focusing on a geodesic (marked in red) that passes through the ${M_r}$ surface, there exist geodesics that intersect it and also pass through the region ${W_r}$ (blue lines), and geodesics do not lie within the ${W_r}$ region (cyan lines). (b) The kinematic picture corresponding to the left figure.}
        \label{fig433}
\end{figure}

We can also understand subregion complexity. As shown in Figure~\ref{42.3a}, as an example, we consider a subregion ${W_r}$ within the bulk, which is defined as the bulk region enclosed by a cutoff surface ${M_r}$. Let us focus on a geodesic (marked in red) that passes through the ${M_r}$ surface. In the quantum circuit perspective, this corresponds to a wire extending from a specific point on the ``kinematic space input board" (see figure~\ref{42.1}). Now consider two other geodesics, one marked in blue that intersects the red geodesic and also passes through the region ${W_r}$, and another marked in cyan that does not lie within the ${W_r}$ region (maybe or may not intersect with the red thread). The corresponding points of these two geodesics on the input board are also present in figure~\ref{42.3b}. Recall that in the sense of surface/state duality, ${M_r}$, as a closed surface, corresponds to a pure state $\Psi_r$. This leads to the complexity meaning of equation (\ref{vol}): it gives the number of quantum gates needed to construct the state $\Psi_r$. The idea is that we need the quantum gates inside ${W_r}$ to construct the state $\Psi_r$, while the quantum gates outside ${W_r}$ are not needed.~\footnote{ Here, we adopt an assumption from surface-state duality implicitly: a point in the interior of ${W_r}$ with no volume corresponds to a state with no internal entanglement, suitable as the initial state for the quantum circuit.} According to our interpretation, the intersection of the blue and red threads in the figure counts as a quantum gate for constructing $\Psi_r$, while the possible intersection of the gray and red threads does not contribute to constructing $\Psi_r$. We can rephrase this from the input board perspective as shown in figure~\ref{42.3b}: let $\partial {W_r}$ (in this case, i.e., ${M_r}$) correspond to a region $\Omega_{\partial {W_r}}$ in kinematic space (in this case, simple a rectangle). For a selected wire $\varsigma_{uv}$ (the red thread), only the quantum gates coupling with the wires located within the region $\Omega_{\partial {W_r}} \cap \diamondsuit_{uv}$ (i.e., the set of blue wires) are needed to construct the quantum state $\Psi_{\partial {W_r}}$ (i.e., $\Psi_r$ in this case). Alternatively, we can equivalently imagine constructing $\Psi_{\partial {W_r}}$ by removing the unnecessary quantum gates (i.e., those corresponding to intersections of the cyan threads with the red threads) from the original quantum circuit (figure~\ref{42.1}). Note that this argument applies to a more general subregion $X$ in the bulk, such as the entanglement wedge $W(A)$ of a boundary subregion $A$, where $\partial {W_r}$ is the union of $A$ and its RT surface.

Interestingly, in a sense, our work provides a more explicit quantum information theory interpretation for the kinematic space, that is, we can understand the kinematic space as a ``circuit input board" that depicts the entanglement structure of holographic spacetime.

\section{Implications of Entanglement Threads}\label{sec4}

In this section, we further discuss the implications embodied by entanglement threads. We will demonstrate that each entanglement thread, in a certain precise sense, characterizes a global entangled state among the multiple qubits it traverses. Hence, the quantum state corresponding to the entire thread configuration can intuitively depict not only the entanglement relationships among the elementary regions on the holographic boundary, but also the entanglement between various surfaces across the entire bulk slice, via those entanglement threads that simultaneously pass through them. This portrayal of entanglement structure can be demonstrated by the fact that the generalized RT formula~ \cite{Miyaji:2015fia, Miyaji:2015yva} within the bulk is satisfied everywhere. We will also discuss the relationship between entanglement threads and the well-known bit threads ~\cite{Freedman:2016zud,Cui:2018dyq,Headrick:2017ucz,Headrick:2022nbe}.

\subsection{Global Perspective on Holographic Entanglement Structure}\label{sec41}

\begin{figure}
     \centering
       \begin{subfigure}[b]{0.46\textwidth}
         \centering
         \includegraphics[width=\textwidth]{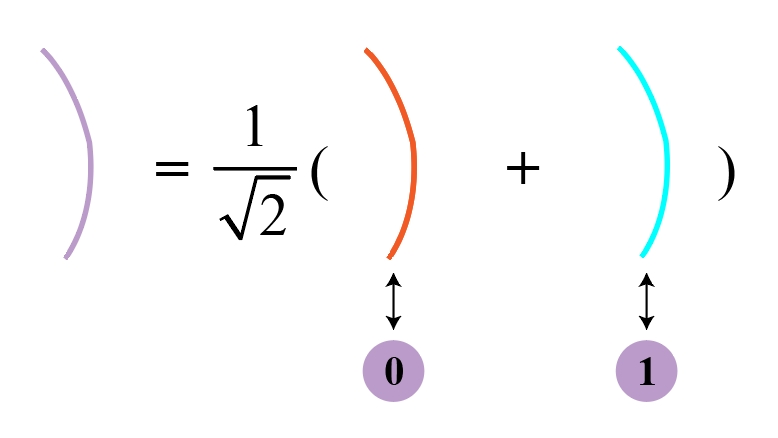}
         \caption{}
         \label{51.1a}
     \end{subfigure}
     \hfill
     \begin{subfigure}[b]{0.45\textwidth}
         \centering
         \includegraphics[width=\textwidth]{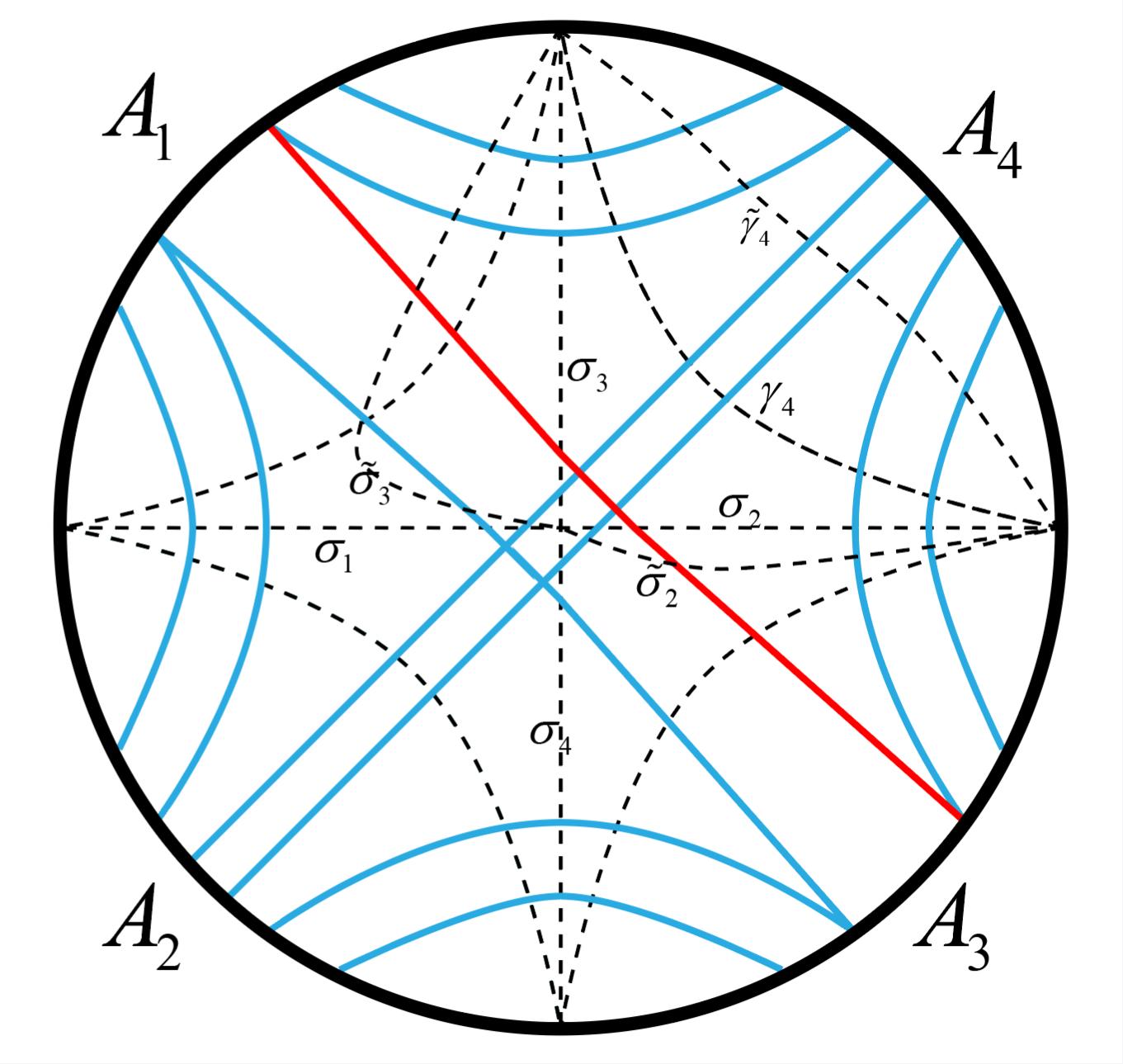}
         \caption{}
         \label{51.1b}
     \end{subfigure}
             \caption{(a) The ``thread-state correspondence’’, first hypothesized in~\cite{Lin:2022flo, Lin:2022agc} based on other motives. (b) A basic quantum state corresponding to the thread configuration, which can be determined by the thread-state correspondence rule. Only a few of the entanglement threads are shown schematically here for illustration.}
        \label{51.1}
\end{figure}

Before proceeding, we offer a nuanced clarification regarding the statement ``the trajectories of entanglement threads are geodesics in the bulk”. When we employ that phrasing, it may appear that we are presupposing the existence of a metric structure of the spacetime background. However, a closer examination reveals that the meaning of an entanglement thread’s ``trajectory” can be understood and defined in a broader sense. Rather than presupposing a metric background, more precisely, we are presupposing the existence of a tensor network endowed with quantum entanglement structure. We begin with a holographic tensor network that can be deformed into a quantum circuit by mapping each basic tensor to a basic quantum gate. In particular, if we arrange the wires according to the organization of kinematic space, then each apparent wire corresponds exactly to an entanglement thread, yielding a canonical quantum circuit. Now, in this canonical circuit, note that the quantum entanglement structure remains a concrete reality, even though the original spatial background and the notion of ``metric” have been deconstructed and dissolved—eliminating talk of ``geodesics”. Nonetheless, we can still discuss the trajectory of an entanglement thread: it can be defined as a partial order determined by the successive sequence of basic quantum gates it passes through—an invariant essence that persists after deforming the original tensor network and dispensing with any spatial metric concept.

Once we understand that the essence of an entanglement thread is a partial order, we realize that we can reverse the process: imagine an initial state equivalent to that obtained by replacing all the non trivial quantum gates in the tensor network model with identity gates. In this case, each wire in the canonical circuit (i.e. each entanglement thread) undergoes only trivial identity evolution. By the mapping state duality in quantum information theory, that is equivalent to stating that within the thread configuration, each entanglement thread corresponds to the following special quantum state:  
\begin{equation}
\left| \zeta \right\rangle = \frac{1}{\sqrt{2}} \bigl( \left| 0_1 0_2 \cdots 0_n \right\rangle + \left| 1_1 1_2 \cdots 1_n \right\rangle \bigr).
\label{thre}
\end{equation}
Here, indices $1$ and $n$ denote the two endpoints of the thread at the boundary of the original space, while the other indices label the qubits in the basic components threaded by the entanglement thread—for example, the internal legs of hexagonal tiles in the HaPPY network. Then the entire thread configuration corresponds to the following quantum state, simply the tensor product of all such thread states:
\begin{equation}
\left| \Psi \right\rangle = \bigotimes_{\text{all }\zeta} \left| \zeta \right\rangle.
\label{coa}
\end{equation}
We take $\left| \Psi \right\rangle$ as an initial state, imagining that we then add quantum gates on top of it to construct the true holographic state. In this sense, the thread configuration state has an advantage not present in other tensor network models: all of them rigidly specify the exact form of the quantum gates, which may deviate from the true holographic state. But the state $\left| \Psi \right\rangle$ occupies a very safe position. It provides partial order information about the true holographic state (i.e. each thread threads bulk lattice points), while avoiding overly restrictive specifications. The point is, as the initial state for preparing the actual holographic state, $\left| \Psi \right\rangle$ is not merely a collection of qudits, but rather qudits endowed with a non-arbitrary partial order structure. One normally expects that only a small set of basic quantum gates might be needed to build the true holographic state; the thread configuration provides the scaffold to place those gates upon.

Even though we have abstracted away from the details of specific quantum gates, $\left| \Psi \right\rangle$ still allows a coarse grained understanding of entanglement in holographic duality. In particular, the thread picture is especially well suited to a global viewpoint on entanglement structure. This globality may reflect an essence of entanglement that transcends circuit details, though it is not yet fully transparent. To appreciate this global perspective, we note that we can rewrite \eqref{thre} in a holistic form: first define each thread $\zeta$ as a superposition of two orthonormal states, formally written as
\begin{equation}
\left| \zeta \right\rangle = \frac{1}{\sqrt{2}} \bigl( \left| \mathrm{red} \right\rangle + \left| \mathrm{blue} \right\rangle \bigr).
\label{rb}
\end{equation}
In this way, the entanglement between the various bulk qudits that are strung together by a single thread can be described as follows (see Figure~\ref{51.1a}): each $ \left| \text{red} \right\rangle $ state represents that all the qudits through which the thread passes are in their own $|0\rangle$ states, and each $ \left| \text{blue} \right\rangle $ state represents that all such qudits are in their own $|1\rangle$ states, i.e.,
\begin{equation}\label{dua}
\begin{array}{l}
\left| \text{red} \right\rangle = \left| 0_1 0_2 \cdots 0_n \right\rangle \\
\left| \text{blue} \right\rangle = \left| 1_1 1_2 \cdots 1_n \right\rangle
\end{array} \end{equation}
Of course, this rewriting actually implies viewing each single wire undergoing trivial dientity evolution in the quantum circuit from a holistic perspective. Interestingly, under this holistic view, we can describe quantum entanglement in terms of the concept of \textit{measurements}. For example, as shown in Figure~\ref{51.1b}, consider a closed surface $ \Sigma = \sigma_2 \cup \sigma_3 \cup \gamma_4 $ in the bulk formed by several extremal surfaces. The thread-state interpretation (\eqref{rb} and \eqref{dua}) generates a quantum state $ \left| \Psi_\Sigma \right\rangle $ associated with this surface. What is the meaning of this quantum state? Consider a new closed surface like $ \tilde{\Sigma} = \tilde{\sigma}_2 \cup \tilde{\sigma}_3 \cup \tilde{\gamma}_4 $ in the figure, which is composed of a set of general bulk surfaces homologous to the extremal surfaces forming the boundary components (such as $ \sigma_2, \sigma_3, \gamma_4 $ in this example). The state $ \left| \Psi_\Sigma \right\rangle $ can be viewed as the entanglement-distilled state associated with the surface.~\footnote{ In other words, it can be hypothesized that $\left| {{\Psi _\Sigma }} \right\rangle$ is asymptotically LOCC equivalent to the true holographic quantum state associated to $\tilde\Sigma$. However, more rigorously, we can only say that the quantum state constructed in this way is an isentropic state of the true holographic state. Simply put, being an isentropic state is a necessary condition for being an asymptotically LOCC-equivalent state~\cite{Bennett:2000}.} It nevertheless represents a coarse-grained characterization of the entanglement relationship between the three basic regions into which the surface $ \tilde{\Sigma} $ is partitioned. $ \left| \Psi_\Sigma \right\rangle $ can be understood under the thread-state interpretation as follows. First, the pure state $ \left| \Psi_\Sigma \right\rangle $ can be formally expressed as a quantum superposition:
\begin{equation}\label{psi}
\begin{array}{l}
\left| \Psi_\Gamma \right\rangle = \sum\limits_Q C_Q \left| Q \right\rangle \\
\left| Q \right\rangle \equiv \left| \sigma_2 \right\rangle \otimes \left| \sigma_3 \right\rangle \otimes \left| \gamma_4 \right\rangle
\end{array} \end{equation}
On the other hand, the state corresponding to the entire thread configuration is given by (\ref{coa}). If we \textit{measure} the thread configuration, it will yield various probability amplitudes for the system to be in various color states. Figure~\ref{51.1b} shows a specific case: suppose we \textit{measure} the thread configuration and find that all the threads starting from $ A_1 $, passing through $ \sigma_3 $ and then $ \sigma_2 $, and reaching $ A_3 $, are in the red state, while all the other threads (including those connecting $ A_1 $ to $ A_3 $ via $ \sigma_1 $ and $ \sigma_4 $) are in the blue state. Then, according to the thread-state correspondence rule, the surface $ \Sigma $ must be \textit{simultaneously} in the following state:
\begin{equation}\label{app}
\left| Q \right\rangle = \left| 1101 \right\rangle \otimes \left| 1011 \right\rangle \otimes \left| 111111 \right\rangle .\end{equation} 
The probability amplitude is only related to the total number of threads passing through $ \Sigma $:
\begin{equation}
N_\Sigma = F_{14} + F_{34} + F_{24} + F_{13}^{\sigma_3 \sigma_2},\end{equation}
that is,
\begin{equation}
C_Q = \left( \frac{1}{\sqrt{2}} \right)^N = \left( \frac{1}{\sqrt{2}} \right)^7.\end{equation}
~\footnote{In other words, the probability of measuring the surface $\Sigma$ to be in the state $ Q $ (\ref{app}) is
\begin{equation}\label{pq}
P_Q = \left| C_Q \right|^2 = \frac{1}{2^N} = \frac{1}{2^7}.\end{equation}}
This is because, for threads that do not pass through $\Sigma$ (e.g., those connecting surface $A_1$ to $A_2$), the measurement that determines their color state does not \textit{affect} whether the qubits on $\Sigma$ are in the 0 or 1 state. On the other hand, each qubit thread that passes through $\Sigma$ provides two possible overall configurations of the two qubits it connects on $\Sigma$: either both are in the 0 state (if the thread is in the red state), or both are in the 1 state (if the thread is in the blue state). Therefore, when there are $N$ threads passing through the surface $\Sigma$, the total number of possible overall states of all qubits on $\Sigma$ is $2^N$, and the probability of obtaining one of these state by measurement is just (\ref{pq}). This kind of analysis can not only determine the pure state corresponding to a closed surface, but also the density matrix of the extremal surface composing the closed surface. For example, consider the density matrix of the extremal surface $\sigma_3$, which can be given by the reduced density matrix relative to $\Sigma$. This means tracing out the complement, i.e., the qubits on $\sigma_2$ and $\gamma_4$. As a result, the endpoint qubits of the threads connecting $\sigma_2$ and $\gamma_4$ have no effect on this process, while for a thread passing through $\sigma_3$, its endpoint qubits (one on $\sigma_3$, one on its complement) entangled in a Bell state, which leads to:
\begin{equation}
\rho_{\sigma_3} = \mathrm{Tr}_{\sigma_2 \cup \gamma_4} \left| \Psi_\Sigma \right\rangle \left\langle \Psi_\Sigma \right| = \left( \frac{1}{2} \left|0\right\rangle \left\langle 0\right| + \frac{1}{2} \left|1\right\rangle \left\langle 1\right| \right)^{(F_{14} + F_{13}^{\sigma_3 \sigma_2} + F_{24}^{\sigma_1 \sigma_3})}.\end{equation}
The exponent here denotes the number of threads with one end on surface $\sigma_3$ and the other on its complement. Clearly, the von Neumann entropy of $\sigma_3$ is exactly equal to this number. Now note that for the surface $\tilde{\Sigma}$, the basic regions composing it are no longer simple extremal surfaces. Therefore, there may be additional threads passing through it, as shown in the figure: there are threads connecting $A_1$ to $A_3$, but passing through surfaces $\sigma_1$ and $\sigma_4$, which intersect surface $\tilde{\sigma}_3$ more than once. However, these extra threads do not directly contribute to the correct entanglement entropy. If we consider the entanglement entropy of $\tilde{\sigma}_3$ relative to the closed surface $\tilde{\Sigma}$, since those threads with both ends located on $\tilde{\sigma}_3$ still have no effect on the partial trace over the complement, 
the von Neumann entropy of $\tilde{\sigma}_3$ is still exactly equal to the number of threads entering the ${{\sigma}_3}$ surface from the $\tilde{\sigma_3}$ surface. We thus obtain a formulation of the generalized Ryu-Takayanagi (RT) formula: The von Neumann entropy of $\tilde{\sigma}_3$ is given by the area of its homologous minimal extremal surface $\sigma_3$, divided by $4{G_N}$~\cite{Miyaji:2015fia, Miyaji:2015yva}.

Although we have only analyzed a simple example, similar discussions can be extended to cases involving more elementary regions and other interior surfaces in the bulk. The idea of associating each thread with a Bell like entangled pair of qudits has been around for a long time. Our point is to explicitly associate each thread explicitly to an overall entangled state among all the qudits it threads. This ensures that no matter which local segment of the thread we examine, the two qudits connected by its two ends are always entangled in a Bell-like state. In this way, we can guarantee that the above discussion holds regardless of which surface in the bulk we choose to analyze. Ultimately, via the full thread configuration state \eqref{coa}, one can intuitively depict not only the entanglement relationships among elementary boundary regions, but also between various surfaces across the entire bulk slice. This portrayal of entanglement structure can be demonstrated by the fact that the generalized RT formula within the bulk is satisfied everywhere.

\subsection{Comparison between Entanglement Threads and Bit Threads}\label{sec42}

\begin{figure}
    \centering
    \includegraphics[scale=0.23]{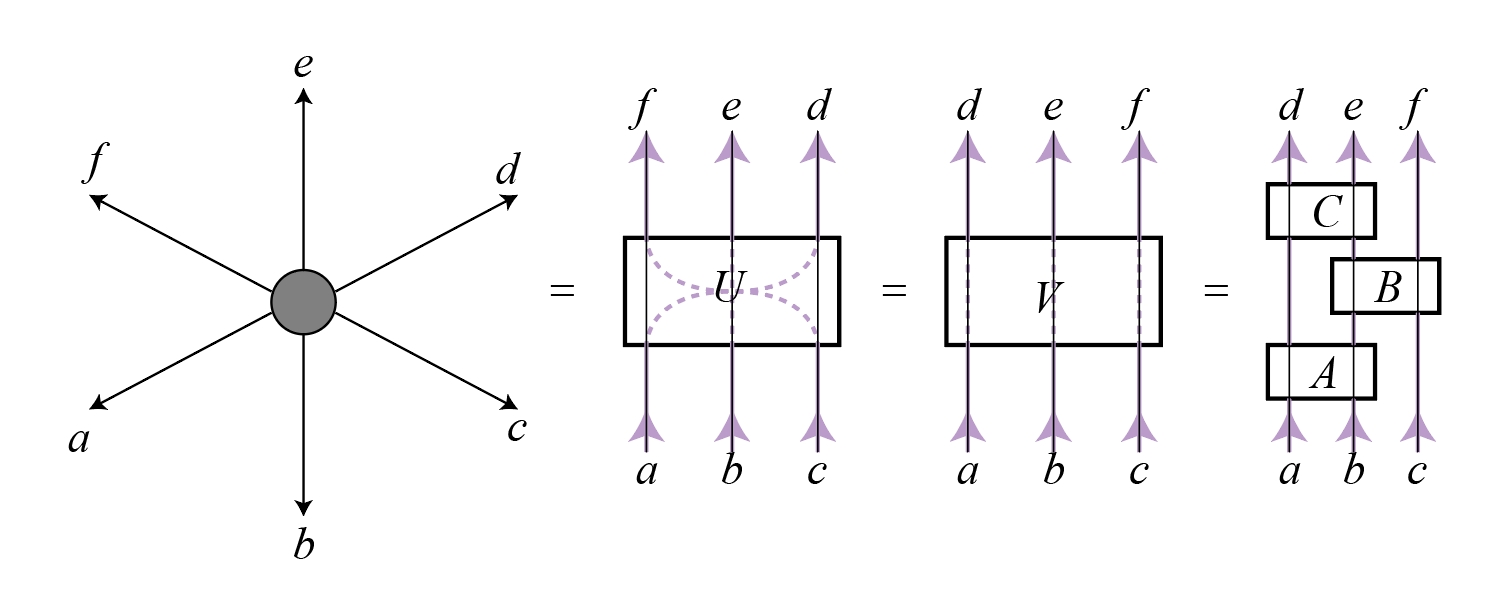}
    \caption{ Due to the conservation of the number of qubits, we have a certain degree of freedom to define the concept of ``apparent wires”(the apparently parallel black lines). Therefore, the configuration of the entanglement threads (purple lines) remains unchanged, while the relatively different locking configurations of bit thread (black lines) may be understood as different choices of apparent‑wire conventions. Note that for $d \geq 5$, a six-rank perfect tensor can be decomposed into a combination of three two-qudit quantum gates~\cite{Pozsgay:2023oyc}. }    \label{52.1}
\end{figure}

As mentioned in the introduction, the thread‑based perspective on entanglement in holographic duality first attracted attention in 2016 with the concept of \textit{bit threads} developed by Headrick et,al. Bit threads are unoriented bulk curves that end on the boundary and subject to the rule that the thread density is less than $\frac{1}{4G_N}$ everywhere~\cite{Freedman:2016zud,Cui:2018dyq,Headrick:2017ucz}.~\footnote{When  one  takes  the  Hodge  dual  of  bit  threads  one  gets  calibrated  geometries,  which  mathematicians (geometers)  use  to  identify  minimal  surfaces.   This  is  a  viewpoint  adopted  in~\cite{Bakhmatov:2017ihw}.}  In particular, this density bound implies that the number of threads passing through a minimal surface $\gamma(A)$, which separates a boundary subregion $A$ from its complement, cannot exceed its area times $\frac{1}{4G_N}$. Therefore, the flux of threads connecting $A$ and its complement, $Flux(A)$, does not exceed this value:
\begin{equation}\label{dens}
Flux(A) \le \frac{1}{4G_N} \, Area(\gamma(A)) .\end{equation}
Using terminology from network flow theory, we say that if inequality (\ref{dens}) is saturated, then a thread configuration \textit{locks the region $A$}. In fact, this bound is tight. For any $A$, there always exists a locking thread configuration such that:
\begin{equation}
Flux_{\text{locking}}(A) = \frac{1}{4G_N} \, Area(\gamma(A)).\end{equation}
This theorem is known as max flow-min cut theorem (see \cite{Headrick:2017ucz} and references therein), that is, the maximal flux of bit threads (over all possible bit thread configurations) through a boundary subregion $A$ is equal to the area of the bulk minimal surface $\gamma \left( A \right)$ homologous to $A$. Therefore, the famous RT formula which relates the entanglement entropy of a boundary subregion $A$ and the area of the bulk minimal extremal surface ${\gamma \left( A \right)}$ homologous to A:
\begin{equation}
 S\left( A \right) = \frac{{Area\left( {\gamma \left( A \right)} \right)}}{{4{G_N}}},\end{equation}
can be expressed in another way, that is, the entropy of a boundary subregion $A$ is proportional to the flux of the locking thread configuration passing through $A$:
\begin{equation}
S(A) = Flux_{\text{locking}}(A).\end{equation}

When the bit threads are required to be locally parallel, one can use the language of $flow$ to describe the behavior of bit threads conveniently in mathematics, that is, using a vector field $\vec v$ to describe the bit threads, just as using the magnetic field $\vec B$ to describe the magnetic field lines. The difference is that for the latter we regard the magnetic field itself as the more fundamental concept, while for the former we consider the threads to be more fundamental. The constraints on the bit threads can then be expressed as the requirements for the flow $\vec v$ as follows,
\begin{equation}
\nabla  \cdot \vec v = 0,\quad \rho \left( {\vec v} \right){\rm{ }} \equiv {\rm{ }}\left| {\vec v} \right| \le \frac{1}{{4{G_N}}}.\end{equation}

For situations involving more than one pair of boundary subregions, the concept of $thread~ bundles$ is also useful. The threads in each thread bundle are required to connect only a specified pair of boundary subregions, while still satisfy the constraints of bit threads. Specifically, one can use a set of vector fields ${\vec v_{ij}}$ to represent each thread bundle connecting the ${A_i}$ region and ${A_j}$ region respectively. The set $V$ of vector fields ${\vec v_{ij}}$ is referred to as a $multiflow$, and each ${\vec v_{ij}}$ is called a $component~ flow$, satisfying (Note that in the present paper we will define ${\vec v_{ij}}$ only with $i < j$ for convenience, which is slightly different from (but equivalent) convention adopted in~\cite{Cui:2018dyq}, where the fields ${\vec v_{ij}}$ were also defined for $i \ge j$, but with the constraint ${{\vec v}_{ji}} =  - {{\vec v}_{ij}}$)
\begin{equation}
\begin{array}{l}
\nabla  \cdot {{\vec v}_{ij}} = 0,\\
\rho (V) \le \frac{1}{{4{G_N}}},\\
\hat n \cdot {{\vec v}_{ij}}{|_{{A_k}}} = 0,\quad ({\rm{for}}\quad k \ne i,j).
\end{array}\end{equation}
It is worth noting that, since in the situation of multiflows, the threads are not necessarily locally parallel, there are various natural ways the density can be defined, and therefore bounded. In particular, \cite{Headrick:2020gyq} discusses how different definitions of thread density can affect the ability to lock a set of boundary regions.

As we have seen in the main text, many properties of our entanglement threads align with the intuition behind bit threads. First, we assume that entanglement threads traverse the bulk continuously without divergence, matching the divergence‑free condition of bit threads. Second, we argued that entanglement threads cannot ``turn back” on RT surfaces, while similarly in an optimal locking thread configuration bit threads do not cross any minimal surface more than once—otherwise they would occupy positions needed by other threads connecting two complementary regions, preventing maximal flux. Most importantly, both frameworks assign entanglement entropy meaning via thread flux. 

However, on the other hand, there are clear differences between the two types of thread configurations. First, bit threads are subject to a density bound, and this bound makes the existence of an optimal bit thread configuration for a given set of regions a nontrivial problem. In simple terms, due to the overly strict constraints on the thread density, it has created an excessive obstacle to the existence of the bit thread configuration that can lock arbitrary number of subregions simultaneously (see the detailed discussion in~\cite{Headrick:2020gyq}). Secondly, the density bound, combined with the max‑flow min‑cut theorem allow more than one thread configurations achieving the same maximal entropy. In other words, the optimal (or locking) bit‑thread configuration can be non-unique. By contrast, our entanglement thread configuration is unique—since we have asserted that each thread’s trajectory is precisely a bulk geodesic. This is the key difference between the two. It is worthwhile adding two remarks that suggest a deeper connection may exist despite these differences. First, in seeking concrete constructions of bit threads, ~\cite{Agon:2018lwq} discovered that geodesics in the bulk naturally serve as the trajectories of bit threads (though, by definition, the bit thread configuration is not unique). Second, ~\cite{Kudler-Flam:2019oru} were the first to point out that the bit thread picture is closely related to the concepts of partial entanglement entropy~\cite{Vidal:2014aal} and conditional mutual information, while in our framework, the entanglement thread flux is also given by half the conditional mutual information.

These observations imply a deeper relationship between the two thread‑based constructions. One possible explanation is that the bit thread configuration captures the distillable portion of entanglement, whereas the entanglement threads define a scaffold that encodes the full entanglement information—on top of which quantum gates may be added to actualize the holographic state. Within our framework, we can attempt to understand the relationship between entanglement thread and bit thread configurations as follows: an entanglement thread configuration is defined by reorganizing the tensor network that encodes the entanglement of a holographic slice into a quantum circuit, whose wires represent entanglement threads. In particular, as in Figure~\ref{42.1}, when the circuit wires are arranged following the kinematic space organization, each apparent wire corresponds exactly to an entanglement thread following a null‑like geodesic trajectory across the slice.

The key point is that in a quantum circuit, due to the conservation of the number of qubits, we have a certain degree of freedom to define the concept of ``apparent wires”(a set of apparently parallel lines in a quantum circuit). A schematic diagram illustrating this freedom is shown in Figure \ref{52.1}. From this point of view, each optimal bit thread configuration corresponds to a choice of apparent wire sets for the quantum circuit, while the canonical one is \ref{42.1}, wherein the entanglement threads in the quantum circuit are perfectly arranged as a series of parallel straight lines, that is to say, each apparent wire corresponds precisely to a geodesic in the holographic geometry.  Thus different locking bit‑thread configurations may be understood as different choices of apparent‑wire conventions. In any case, more subtle relations between entanglement threads and bit threads may be a fruitful topic for further exploration.

\section{Conclusion and Discussion}\label{sec5}

This paper has depicted a thread picture using entanglement threads to characterize entanglement structures in holographic duality. This framework is based on two important insights. The first is a simple idea or philosophy: ``represent (holographic) quantum entanglement with connecting lines''. The second is that the entanglement thread picture naturally and universally emerges in tensor network models of holographic duality. For instance, the partial order structure associated with the HaPPY tensor network naturally gives rise to local directed arrows. By gluing together these local arrows, one obtains global entanglement threads. By abstracting these concepts from specific tensor network models, we can imagine entanglement threads traversing through spatial slices of the holographic bulk, with their endpoints anchored on the holographic boundary. This is very similar to the earlier concept of bit threads. However, there exist important distinctions between the two. 

We argued that the specific trajectories of entanglement threads in the bulk are exact geodesics. This naturally connects to the concept of kinematic space, which is defined as the collection of all bulk geodesics anchored to boundary points. As a result, kinematic space not only provides mathematical tools for computing the flux of entanglement threads, but also offers a natural way to organize qudits, enabling the construction of a canonical quantum circuit in which each apparent wire precisely corresponds to an entanglement thread. In other words, by interpreting entanglement threads as a set of predefined wires, one can generate holographic quantum states simply by continuously inserting basic quantum gates. We have also tested this idea using the concept of holographic complexity. Combining the concepts of entanglement threads and kinematic space, a elegant circuit interpretation for the holographic complexity is provided. In a certain sense, our work provides a clearer quantum information-theoretic meaning for kinematic space, namely, we can view kinematic space as a ``circuit board'' that encodes the entanglement structure of spacetime.

As emphasized in Section~\ref{sec4}, the concept of entanglement thread trajectory does not require a pre-existing spacetime background with a metric; rather, it can essentially be defined as a partial order specifying the sequence in which it passes through a set of basic quantum gates. During the deformation of the original tensor network, although the metric concept of space is eliminated, this partial order is preserved. We can also assign to each entanglement thread a thread-state correspondence interpretation, which expresses the global entanglement among the qudits it traverses. Ultimately, through the quantum state of the entire thread configuration, not only the entanglement among subregions on the holographic boundary, but also the entanglement relations between different surfaces on the bulk slice are intuitively characterized by the threads that simultaneously pass through them.

The idea of entanglement threads offers a framework to explore many interesting questions. For instance, we can further explore more types of holographic geometries, such as wormholes~\cite{Banados:1992gq} and the dynamical aspects of perturbed AdS spaces. On the other hand, what we studied here is a ``thread game'' for spacelike geodesics. An interesting idea is to consider a similar thread game for timelike geodesics. We conjecture this may be related to pseudo-entropy~\cite{Nakata:2020luh} (since it is a kind of timelike entanglement entropy~\cite{Doi:2023zaf}) or complexity~\cite{Susskind:2014rva, Stanford:2014jda, Brown:2015bva, Brown:2015lvg} (inspired by the Lorentzian bit threads proposed in~\cite{Pedraza:2021mkh, Pedraza:2021fgp, Caceres:2023ziv }). In~\cite{Lin:2023hzs, Lin:2022aqf}(see also\cite{Wen:2024uwr}), we have tried to use the thread picture to understand holographic BCFT and the island phenomenon, and found interesting connections between the island and perfect entanglement. It is also worthwhile to further understand those results from the perspective of quantum circuits and kinematic space developed in this paper.

\newpage
\begin{appendix}
\section*{Appendix}

\section{  Review of Kinematic Space}\label{appa}

\subsection{ Kinematic Space and the Crofton Form}

\begin{figure}
     \centering
     \begin{subfigure}[b]{0.35\textwidth}
         \centering
         \includegraphics[width=\textwidth]{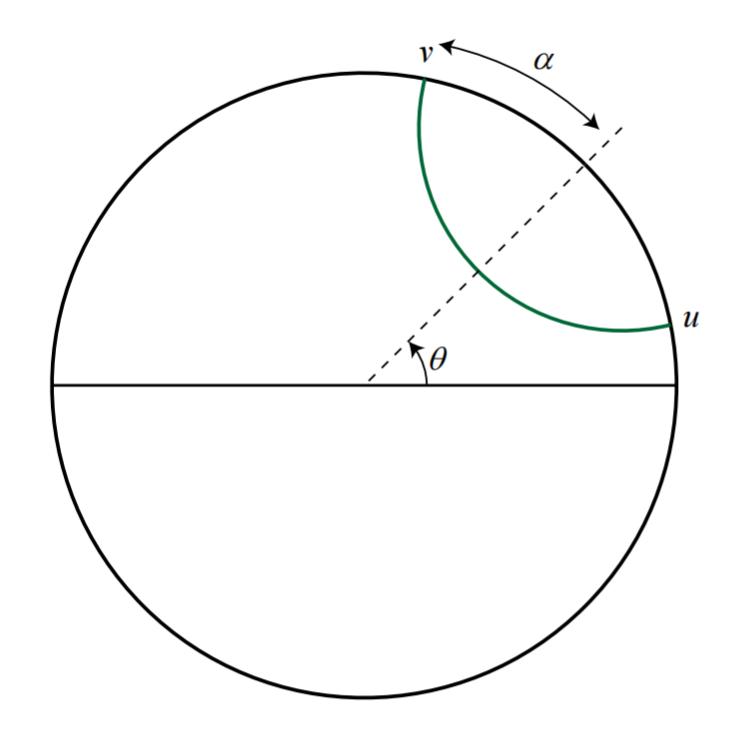}
         \caption{}
         \label{figa1a}
     \end{subfigure}
     \hfill
     \begin{subfigure}[b]{0.34\textwidth}
         \centering
         \includegraphics[width=\textwidth]{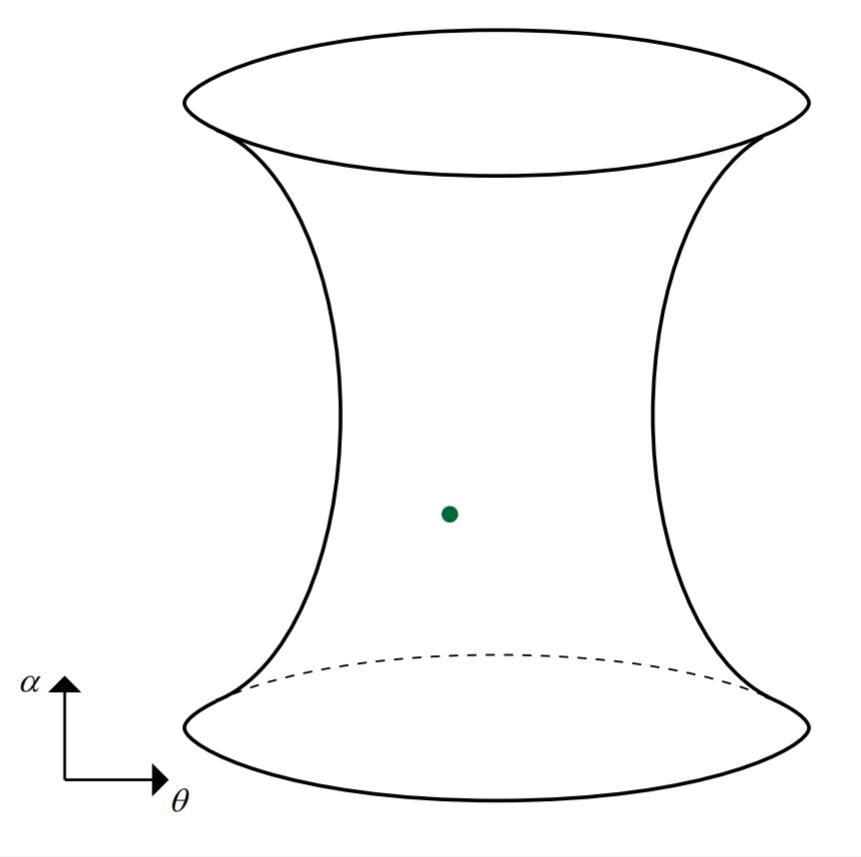}
         \caption{}
         \label{figa1b}
     \end{subfigure}
     \caption{ (a) $\alpha$ is the half-opening angle of the geodesic, and $\theta$ marks the center point of the corresponding boundary subregion. (b) Each point in kinematic space $K$ represents a geodesic in the original space $N$. }
\label{figa1}
\end{figure}

The concept of kinematic space~\cite{Czech:2015kbp,Czech:2015qta} originates from the application of the Crofton formula in integral geometry\cite{integral} and its extension in holographic geometry. The Crofton formula roughly states that the length of a curve is measured by the number of geodesics intersecting it. In the context of holographic duality, consider a static slice of a holographic geometry. Without loss of generality, we take an equal-time slice $ N $ of $\text{AdS}_3$ space, with the metric given by 
\begin{equation}\label{ads3}
ds^2 = l_{\text{AdS}}^2(d\rho^2 + \sinh^2\rho \, d\phi^2) .\end{equation}
Each geodesic can be parametrized as:
\begin{equation}
\cos(\alpha) = \tanh(\rho)\cos(\phi - \theta),\quad \alpha \in (0,\pi),\quad \theta \in S^1,\end{equation}
as shown in Figure \ref{figa1}, here $\alpha$ intuitively corresponds to half the opening angle of the geodesic, while $\theta$ labels the center of the boundary interval corresponding to the geodesic on the boundary $ M = \partial N = S^1 $. Each complete oriented geodesic is uniquely specified by a pair $(\theta, \alpha)$. Note that reversing the direction of the geodesic $(\theta, \alpha)$ corresponds to $(\theta + \pi, \pi - \alpha)$.

As illustrated in Figure \ref{figa1}, by mapping each geodesic's label $(\theta, \alpha)$ to a point $(\theta, \alpha)$ in a new space $ K $, the counting of geodesics is translated into volume measurement in the new space. This new space $ K $, dual to the original space $ N $, is called the kinematic space. More specifically, it was proposed that the length of a curve $ e $ in $ N $ is given by:
\begin{equation}\label{leng}
\text{Length}(e) = \frac{1}{4} \int_K \omega(\theta, \alpha) \, n_e(\theta, \alpha) ,\end{equation}
where $ n_e(\theta, \alpha) $ is the number of intersections between the geodesic $(\theta, \alpha)$ and the curve $ e $, and the integration measure $ \omega $ is physically interpreted as the ``geodesic density function’’. It is the volume form in kinematic space $ K $, also known as the Crofton form. Specifically, for a static slice of $\text{AdS}_3$, the Crofton form is given by:
\begin{equation}\label{w1}
\omega(\theta, \alpha) = -\frac{1}{\sin^2\alpha} \, d\alpha \wedge d\theta .\end{equation}
In the literature, another set of parameters is also commonly used:
\begin{equation}\label{uv}
u = \theta - \alpha,\quad v = \theta + \alpha .\end{equation}
Under this change of variables, Eq.~(\ref{w1}) becomes:
\begin{equation}\label{w2}
\omega(u, v) = \frac{1}{2 \sin^2\left(\frac{v - u}{2}\right)} \, du \wedge dv .\end{equation}
In other words, the kinematic space $ K $ is a manifold with the metric:
\begin{equation}\label{ds}
ds^2 = \frac{1}{2 \sin^2\left(\frac{v - u}{2}\right)} \, du \, dv ,\end{equation}
which is, in fact, a $\text{dS}_2$ (two-dimensional de Sitter) space.

For more general static slices $ N $ of holographic geometries, the Crofton measure $ \omega $ can be derived as follows: first obtain the length $ \lambda(u,v) $ of a geodesic in $ N $ parametrized by $(u, v)$ (note that $u, v$ are exactly the boundary spatial coordinates of the two endpoints of the geodesic on the boundary $ S^1 $), then
\begin{equation}\label{w3}
\omega = \frac{\partial^2 \lambda(u,v)}{\partial u \, \partial v} \, du \wedge dv .\end{equation}

\subsection{ Point-Curves and Conditional Mutual Information}

\begin{figure}
     \centering
       \begin{subfigure}[b]{0.46\textwidth}
         \centering
         \includegraphics[width=\textwidth]{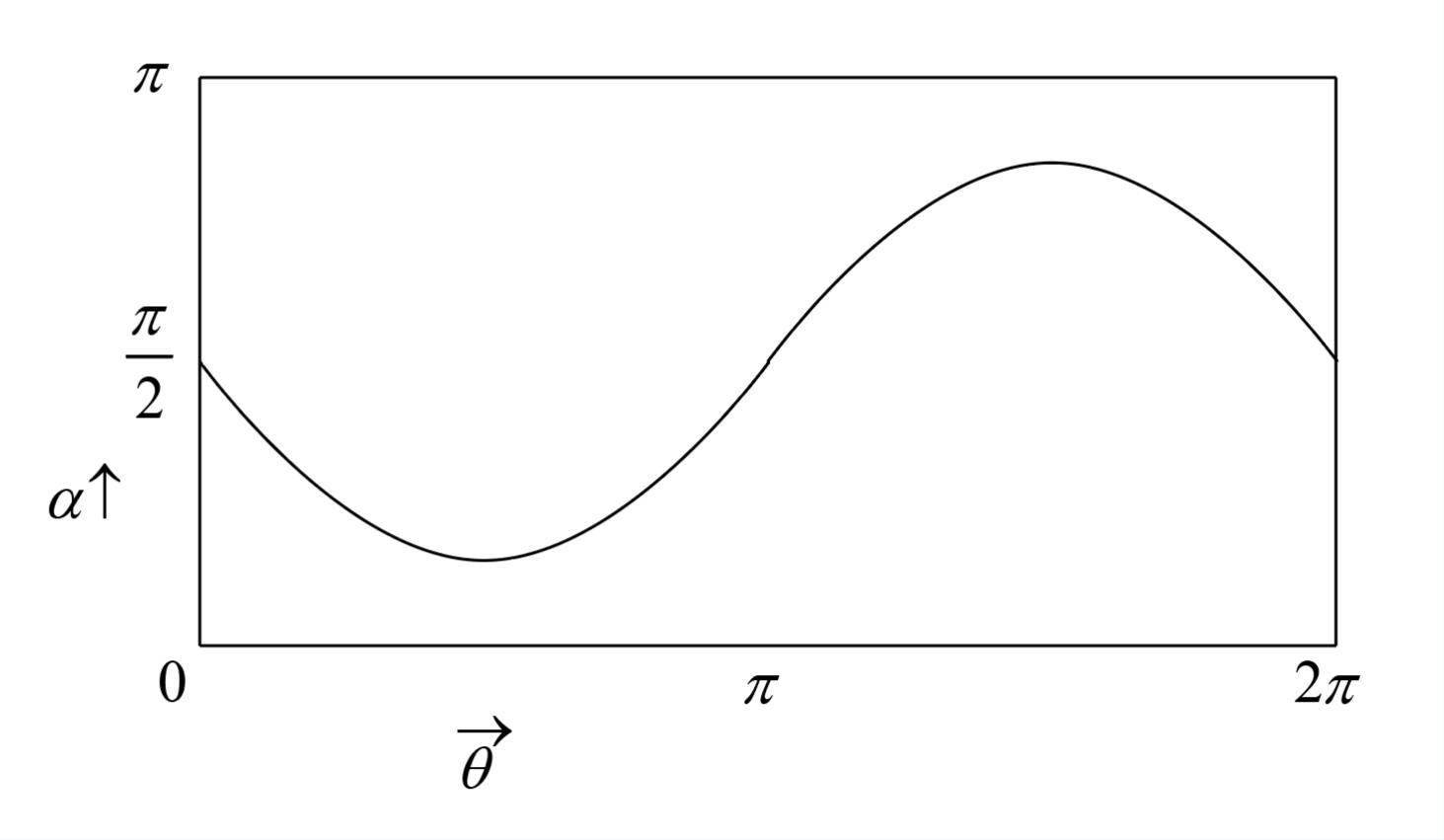}
         \caption{}
         \label{figa2a}
     \end{subfigure}
     \hfill
     \begin{subfigure}[b]{0.45\textwidth}
         \centering
         \includegraphics[width=\textwidth]{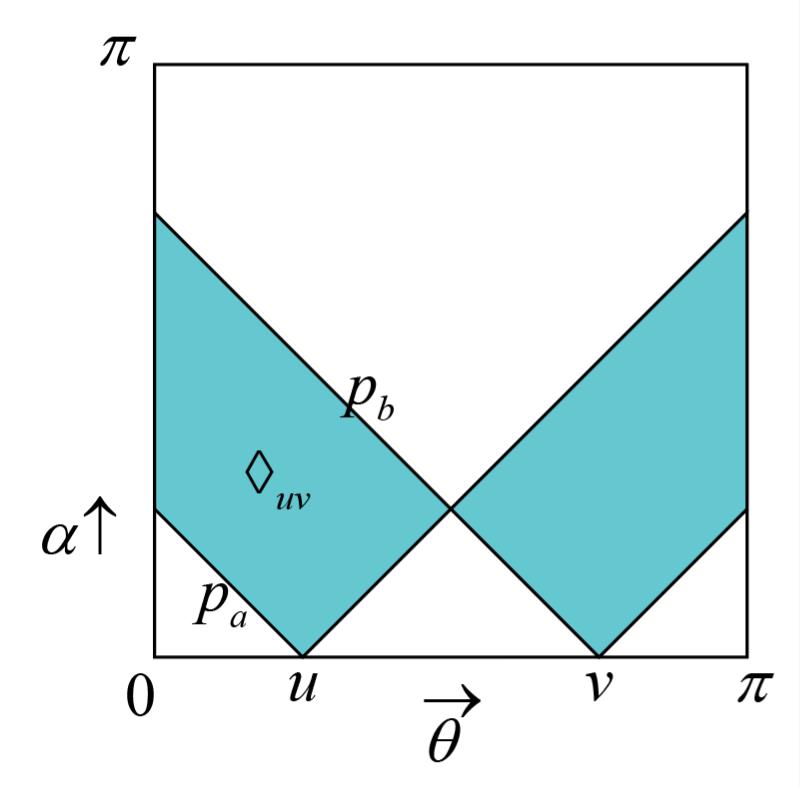}
         \caption{}
         \label{figa2b}
     \end{subfigure}
             \caption{(a) A point-curve in $K$ corresponding to a point in $N$. (b)  Computing the geodesic distance between two boundary points in $N$. Note that half of $K$ space already contains all the geodesic information in $N$ space. }
        \label{figa2}
\end{figure}

Let us illustrate how the language of kinematic space can be used to characterize the geometric information in the original space $ N $, via an example of computing the length of a geodesic in $ N $. A point $ a $ in $ N $ can be mapped to $ K $ in the following way: find all geodesics in $ N $ that pass through the point $ a $. Each such geodesic corresponds to a point in $ K $, so the set of all such geodesics forms a trajectory in kinematic space $ K $, denoted $ p_a $, which is called a point-curve. In particular, for pure $\text{AdS}_3$ (and also BTZ black holes), it can be shown that the point-curves are spacelike geodesics in $ K $ \cite{Czech:2014ppa}. In the pure $\text{AdS}_3$ case, the point $(\rho, \phi)$ in the coordinate system (\ref{ads3}) corresponds to a point-curve in $ K $ under coordinates $(\theta, \alpha)$ given explicitly by~\cite{Czech:2014ppa, Czech:2015qta}:
\begin{equation}
\alpha(\theta) = \cos^{-1}(\tanh \rho \cdot \cos(\theta - \phi)).\end{equation}
Now, using the mathematical language of kinematic space, computing the length of the geodesic ${e_{ab}}$ between two bulk points $a$ and $b$ in the original space $N$ becomes straightforward: we respectively draw the point curves $p_a$ and $p_b$ corresponding to $a$ and $b$ in kinematic space $K$, which enclose a region denoted by ${p_a}\Delta {p_b}$. Next, we can apply formula (\ref{leng}). First, we specify the integration region as ${p_a}\Delta {p_b}$; second, noting that since ${e_{ab}}$ is itself a geodesic, other geodesics can cross it at most once, the formula (\ref{leng}) simplifies to:
\begin{equation}\label{eab}
Length({e_{ab}}) = \frac{1}{4}\int\limits_{{p_a}\Delta {p_b}} \omega .\end{equation}
In particular, let us consider the case when both $a$ and $b$ lie on the boundary $\partial N = M$. Note that since $\partial K = \partial N = M$, the corresponding point curve of a boundary point $a$ in $N$ in kinematic space will exactly be the ``light cone" emanating from the same boundary point $a$ in $K$, as shown in Figure \ref{figa2}. Now a very interesting result emerges: according to formula (\ref{eab}), the length of the geodesic between two boundary points $a$ and $b$ in the original space $N$ is given by the volume of the blue region in $K$ space shown in Figure \ref{figa2b}. Let the coordinates of points $a$ and $b$ on the boundary be $u$ and $v$, and denote the geodesic length between them as $\lambda(u,v)$. Let the blue region be denoted by ${\diamondsuit_{uv}}$, then we obtain the formula for computing geodesic lengths using the language of kinematic space:
\begin{equation}\label{lamb}
\lambda(u,v) = \frac{1}{4}\int_{{\diamondsuit_{uv}}} \omega .\end{equation}
However, from the RT formula, we know that the left-hand side of (\ref{lamb}) precisely gives the entropy of the segment $R = [u,v]$, namely $S(R) = S(u,v)$, that is:
\begin{equation}
S(u,v) = \frac{\lambda(u,v)}{4G_N}  .\end{equation}
Thus, a close connection arises between the kinematic space and the concept of entanglement entropy! By renormalizing the Crofton form, define:
\begin{equation}
\tilde{\omega} = \frac{\omega}{4G_N},\end{equation}
then we obtain:
\begin{equation}\label{suv}
S(u,v) = \frac{1}{4}\int_{{\diamondsuit_{uv}}} \tilde{\omega},\end{equation} 
that is, the entanglement entropy of a (connected) subregion $R$ in the boundary CFT can exactly be represented as the volume of a corresponding region ${\diamondsuit_{uv}}$ in the kinematic space. The region ${\diamondsuit_{uv}}$, called the causal diamond, is defined as the diamond-shaped region in kinematic space bounded by the light rays emanating from the two endpoints $u$ and $v$ of $R$ (and its complement $\bar{R}$).

\begin{figure}
    \centering
    \includegraphics[scale=0.37]{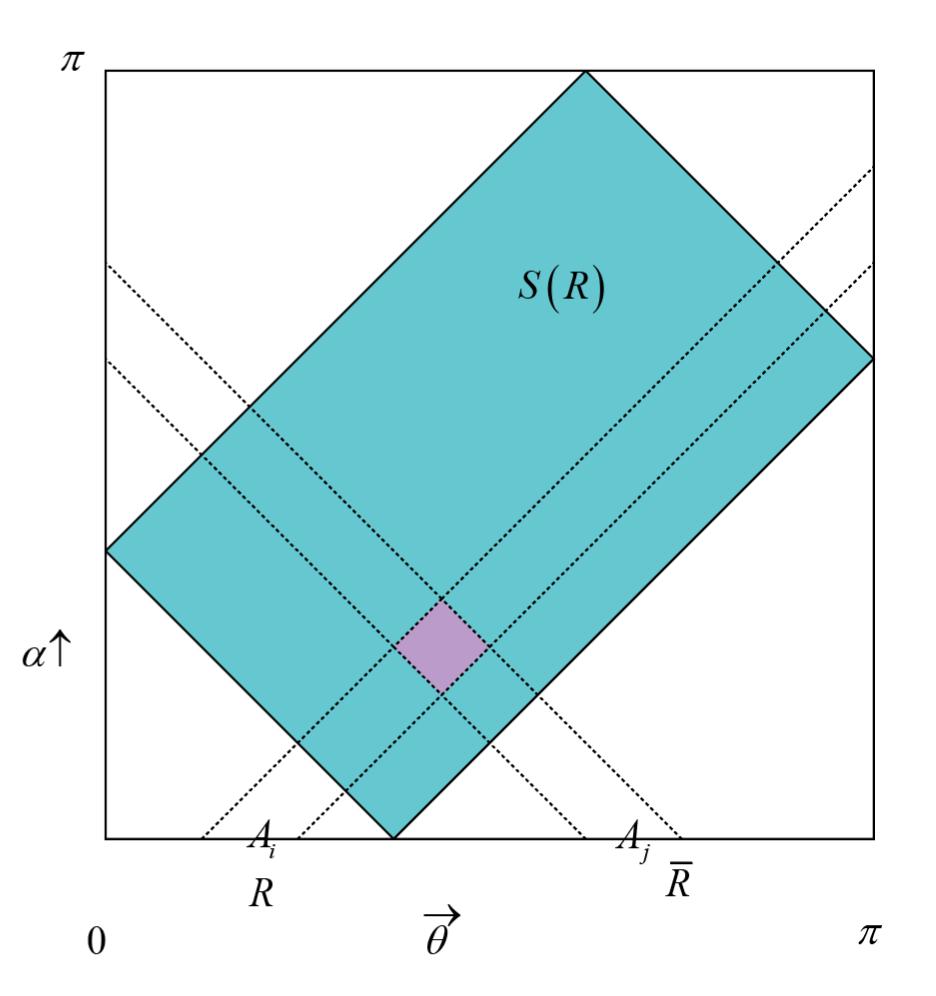}
    \caption{The half-qCMI has a direct and intuitive geometric interpretation in kinematic space. $\frac{1}{2}I(A_i,A_j\mid L)$ corresponds to the volume of purple diamond region. The coordinate $\theta$ has been chosen such that one of the endpoints of $R$ is exactly the origin. }
    \label{figa3}
\end{figure}

The half-conditional mutual information also has an intuitive graphical representation in kinematic space. As shown in the figure~\ref{figa3}, consider two fundamental regions ${A_i}$ and ${A_j}$ of the boundary system, with the intermediate region denoted by $L$. Then, the half-conditional mutual information $\frac{1}{2}I(A_i,A_j\mid L)$ (see (\ref{half})) is exactly given by the volume of a diamond region ${\diamondsuit_{A_i,A_j\mid L}}$ in kinematic space. It is defined as the region enclosed by intersecting light rays emanating from the endpoints of ${A_i}$ and ${A_j}$ at the boundary of kinematic space. It can be verified that:
\begin{equation}
\frac{1}{2}I(A_i,A_j\mid L) = \frac{1}{4} \int_{{\diamondsuit_{A_i,A_j\mid L}}} \tilde{\omega}.\end{equation}
In our setup, we usually take the fundamental regions ${A_i}$ and ${A_j}$ to be very small, so this formula can also be directly seen from (\ref{w3}).

In the context of our study of holographic entanglement threads, since we identify each geodesic as an  entanglement thread, this intuitive correspondence of half-conditional mutual information is consistent with our equation (\ref{equ}). As shown in the figure, suppose the fundamental regions ${A_i}$ and ${A_j}$ respectively belong to a bipartition of the boundary system, with ${A_i} \subseteq R,\;{A_j} \subseteq \bar R$. The figure also shows the diamond region ${\diamondsuit_R}$ used to compute the entanglement entropy between $R$ and $\bar R$. Obviously, by decomposing $R$ into a series of $A_i$ and $\bar R$ into a series of $A_j$, the volume of ${\diamondsuit_R}$ can exactly be partitioned into the sum of the volumes of the corresponding ${\diamondsuit_{A_i,A_j \mid L}}$. This precisely expresses that the number of threads connecting $R$ and $\bar{R}$ can be decomposed into the contributions of thread bundles connecting each pair of elementary regions ${A_i}$ and ${A_j}$, with a flux value of $\frac{1}{2}I(A_i,A_j \mid L)$. A preliminary discussion can be seen in~\cite{Lin:2022flo}.

\end{appendix}

\newpage{\pagestyle{empty}\cleardoublepage}

\end{document}